\documentclass[11pt,a4paper]{article}
\pdfoutput=1
\usepackage{jheppub}
\usepackage{subfigure}
\def\hbar{\hspace{0pt}\raisebox{1pt}{$-$} \hspace{-7pt} h}

\def\5{\overline 5}

\newcommand{\ba}{\begin{eqnarray}}
\newcommand{\ea}{\end{eqnarray}}
\newcommand{\no}{\nonumber}
\newcommand{\be}{\begin{equation}}
\newcommand{\ee}{\end{equation}}
\newcommand{\bea}{\begin{eqnarray}}
\newcommand{\eea}{\end{eqnarray}}

\title{Asymmetric WIMP dark matter}
\date{\today
}
\author{
Michael L. Graesser, Ian M. Shoemaker, and Luca Vecchi}

\affiliation{Theoretical Division T-2, Los Alamos National Laboratory \\ Los Alamos, NM 87545, USA}
\emailAdd{mgraesser@lanl.gov}
\emailAdd{ianshoe@lanl.gov}
\emailAdd{vecchi@lanl.gov}

\abstract{

In existing dark matter models with global symmetries the relic abundance of dark matter is
either equal to that of anti-dark matter (thermal WIMP), or vastly larger,
with essentially no remaining anti-dark matter (asymmetric dark matter). 
By exploring the consequences of a primordial asymmetry on the coupled dark matter and anti-dark matter Boltzmann equations we
find large regions of parameter space that interpolate between these two
extremes. Interestingly, this new {\it asymmetric WIMP} framework can accommodate a wide range of dark matter masses and annihilation cross sections. %In fact, the dark matter relic abundance is accounted for by cross sections equal or larger than the thermal WIMP value, and by dark matter masses not exceeding an upper bound determined by the mass of a totally asymmetric candidate. 
 The present-day dark matter population is typically asymmetric, but only weakly so, such that indirect signals of dark matter annihilation are not completely suppressed.  We apply our results to existing models, noting that upcoming direct detection experiments will constrain a large region of the relevant parameter space.

}

\keywords{Beyond Standard Model; Supersymmetry Phenomenology; Cosmology of Theories beyond the SM.
}
\begin{document}
\maketitle

\section{Introduction}

Andrei Sakharov~\cite{Sak} pointed out that a primordial particle number asymmetry is produced given three ingredients: symmetry violation, C and CP violation, as well as a departure from thermal equilibrium. The observed baryon asymmetry suggests that these conditions have been met in the history of the Universe. Since global symmetry violation is expected to occur in the early Universe, we expect the existence of primordial asymmetries to be rather generic.

The dominant contributions to the observed matter density in the Universe are due to dark matter and baryons. Their density fractions are remarkably well determined from cosmic microwave background data from WMAP7~\cite{WMAP7}:
\ba\label{Omegas}
\Omega_{DM} h^{2} =0.1109 \pm 0.0056\quad\quad\quad\Omega_{B}h^{2} =0.02258^{+0.00057}_{-0.00056}.
\ea
Yet, very little remains known about the nature of dark matter (DM). In most theoretical efforts, the problems of dark matter and baryogenesis are treated separately, with the implicit assumption that the comparable densities of these two types of matter is simply a coincidence. However, the fact that the abundances of baryons and DM are just a factor $\sim5$ apart may be an indication of a common underlying origin. The paradigm of asymmetric dark matter (ADM) has been suggested as a way of linking the asymmetries, and thus the abundances in the dark and visible sectors; see~\cite{ADM} and references therein~\footnote{There are also many other models seeking to link the abundances of baryons and DM~\cite{Dodelson,Kaplan,Thomas,enqvist,Kitano,GravQ,Hylogenesis,Allahverdi,gaugedb,mcd,gu}.}.

In ADM two distinct scenarios are typically considered~\cite{nussinov,Barr:1990ca,hooper,Darkogenesis, adm16,adm15,adm14,adm13,adm12,adm11,adm10,adm9,adm8,adm7,adm6,adm5,adm4,adm3,adm2,adm1,Falko,Xogenesis,f1,f2,f3,Kohri:2009yn,Agashe:2004bm,Farrar:2004qy,Khlopov,Wise}.  In one case, a primordial asymmetry in one sector is transferred to the other sector. Here the primordial DM ($\eta_{DM}$) and baryonic ($\eta_B$) asymmetries generally satisfy either $\eta_{DM}\sim\eta_B$ or $\eta_{DM}\ll\eta_B$ depending on whether the transfer mechanism decouples when the DM is relativistic or non-relativistic%\cite{Barr:1990ca,f2,f3,adm9,adm15,Xogenesis}
, respectively.  Whereas if both asymmetries are generated by the same physical process then $\eta_{DM}\gg\eta_B$ is also possible. Now, if, in analogy with the visible sector, the present-day DM population is totally asymmetric, one can explain the coincidence expressed in~(\ref{Omegas}) by \emph{tuning} the DM mass to the value
\ba\label{mDM}
m_{ADM}=\frac{\Omega_{DM}}{\Omega_B}\frac{\eta_B}{\eta_{DM}}m_p,
\ea
with $m_p$ denoting the proton mass. 

The final DM abundance is rendered asymmetric by removing the conventional symmetric component that arises from thermal freeze-out. In the existing literature, this is typically achieved by the introduction of either a strong coupling, in analogy with QCD for the visible sector, or new light states. Both ingredients effectively induce an annihilation cross section sufficiently large to suppress the thermal symmetric component.  Our results alleviate this hurdle to ADM model building and bolsters the case for their further study.  

Here we consider the coupled evolution of the symmetric and asymmetric populations via the Boltzmann equations and quantify how large the annihilation cross section needs to be in order for the present DM densities to be asymmetric. We will see that the symmetric population depends \emph{exponentially} on the annihilation cross section, and hence that a weak scale force typically suffices to remove the symmetric component. In this sense, the weakly interacting massive particle (WIMP) thermal freeze-out scenario is not incompatible with the paradigm of ADM.

We then discuss the implications for ADM and point out that the spectrum of possible ADM scenarios is much richer than previously thought. In particular, we claim that most weakly coupled extensions of the standard model are likely to interpolate between the extreme cases of purely asymmetric DM and symmetric DM.  

In this {\it asymmetric WIMP} scenario the present-day DM abundance is determined by a combination of its thermal annihilation cross section, its mass, and the primordial asymmetry.  This relation among the short-distance and primordial parameters generalizes what is required in the extremely asymmetric or symmetric scenarios.  In fact, for an asymmetric WIMP the total dark matter abundance is allowed to depend on a new observable, namely the ratio of the anti-dark matter to dark matter number densities.  In contrast, in the asymmetric scenarios considered so far this ratio is fixed to be zero, and consequently the present abundance is set by fixing the mass as in~(\ref{mDM}). Likewise, in the symmetric scenario  the dark matter and anti-dark matter abundances are assumed to be exactly equal,  and the present abundance is then set by fixing the annihilation cross section.  For asymmetric WIMP dark matter, these two scenarios are recovered as limiting cases of either small or large present-day anti-DM particle abundances.  

We then apply our results to some existing models of ADM transfer operators \cite{ADM}, and investigate more generally the utility of weak-scale suppressed higher dimensional operators. This is motivated by our findings that 
an asymmetric WIMP needs an annihilation cross-section only a few times larger than a picobarn to obtain the correct dark matter abundance.  As an illustration, for the Higgs portal we find that requiring the correct annihilation cross-section leads to spin-independent direct detection rates at currently observable levels. The next round of direct detection experiments can cover a substantial part of the parameter space. 

The outline of the paper is as follows. In Sec.~\ref{origin} we solve the Boltzmann equations for a generic species in the presence of a primordial particle/anti-particle asymmetry.  In Sec.~\ref{adm} we specialize our results to the case in which the species is the DM, and discuss the implications of the asymmetric WIMP.  In Sec.~\ref{adm:app} we apply our results to existing scenarios of ADM and determine the limits imposed by direct detection experiments.  We summarize our results in Sec.~\ref{conc}.  

In Appendix~\ref{CT} we present accurate results for $2\rightarrow n$ collision terms in the approximation that the incoming particles are non-relativistic on average and the final state threshold is much larger than the masses of the incoming particles. In Appendix~\ref{colapps} we apply the results of the previous Appendix to the specific transfer operators considered in the text. Finally in Appendix~\ref{appDD} we compute the elastic scattering cross section for direct detection arising from the Higgs portal operator.

\section{On the origin of asymmetric species}
\label{origin}
We begin with an analysis of the effect of a primordial particle-antiparticle asymmetry on a generic species $X$ of mass $m$. Previous work on the relic abundance in the presence of an asymmetry appears in \cite{GS,ST}.

We assume that the particle $X$ is \emph{not} self-conjugate~\footnote{This is certainly the case if $X$ carries a $U(1)_X$ global number, but our results apply more generally.}, and that a particle/anti-particle asymmetry in the $X$-number is generated at high temperatures. As the Universe expands, the number violating effects decouple at a temperature $T_{D}$ and the asymmetry is frozen in for $T< T_{D}$. At this stage the number density of particles ($n^+$) and antiparticles ($n^-$) is controlled by a set of coupled Boltzmann equations. Under reasonable assumptions~\footnote{In writing~(\ref{Boltz}) we assumed that i) our particle species is in a bath of particles in thermal equilibrium, ii) that no mass degeneracy between the two sectors is present, iii) that the dominant process changing the $n^\pm$ densities is annihilation, and iv) that the annihilation process occurs far from a resonant threshold~\cite{GS1}.} these reduce to:
\ba\label{Boltz}
\frac{dn^\pm}{dt}+3Hn^\pm=-\langle\sigma_{\textrm{ann}}v\rangle\left(n^+n^--n_{eq}^+n_{eq}^-\right),
\ea
where $\langle\sigma_{\textrm{\small{ann}}}v\rangle\equiv\sigma_0 (T/m)^n$ is the thermally averaged annihilation cross section times the relative particle velocity. In the absence of significant entropy production it is convenient to introduce the quantity $Y^\pm\equiv n^\pm/s$, where $s$ denotes the total entropy density. In terms of this quantity, the particle-antiparticle asymmetry can be expressed by
\ba
\eta\equiv Y^+-Y^-.
\ea
Without loss of generality, we conventionally define \emph{particles} to be more abundant than \emph{anti-particles} such that $\eta \ge 0$. Notice that, as anticipated, $\eta$ stays constant in the thermal evolution described by~(\ref{Boltz}).

The present densities $Y^\pm_\infty$ turn out to be strong functions of the primordial asymmetry $\eta$ \emph{and} the annihilation cross section $\sigma_0$. We distinguish between two limiting scenarios: 
\begin{itemize}
\item In the first regime the dynamics can be considered ``strong" and the final abundance is dominated by particles over antiparticles. In this case $Y^-_\infty\ll Y^+_\infty\simeq\eta$. This is typical of most baryogenesis schemes if $X$ is identified with the ordinary baryons. 
\item In the second regime the dynamics is ``weak" and the final abundances of particles and antiparticles are comparable, and separately much bigger than the asymmetry. In this case $Y^-_\infty\simeq Y^+_\infty\gg\eta$. This is generally assumed in the standard thermal WIMP scenario if $X$ is identified with the dark matter.
\end{itemize}

In the remainder of this section we will present a careful study of the Boltzmann equations~(\ref{Boltz}) in the presence of a primordial asymmetry, and provide a quantitative assessment of what ``strong" and ``weak" dynamics mean (see Sec.~\ref{analyticsol}). We will see that there is a broad range in parameter space between the two extremes discussed above in which the final abundances are comparable, namely
\ba\label{intermediate}
Y^-_\infty\sim Y^+_\infty\sim\eta.
\ea

\subsection{The fractional asymmetry}

In order to solve the coupled system~(\ref{Boltz}) it is useful to introduce the quantity
\ba
r=\frac{n^-}{n^+}=\frac{Y^-}{Y^+}.
\ea
Because the asymmetry $\eta$ is conserved, knowledge of the above quantity 
suffices to determine the relative abundances at any time via~\footnote{As anticipated, the following discussion does not depend on our convention that particles are more abundant that anti-particles. In particular, the physics is invariant under the transformation $\eta\rightarrow-\eta$, $r\rightarrow1/r$.}   
\ba\label{YY}
Y^+=\eta\,\frac{1}{1-r}\quad\quad Y^-=\eta\,\frac{r}{1-r}. 
\ea
Notice that by definition $r$ satisfies $0\leq r\leq1$. 

We refer to $r$ as the \emph{fractional asymmetry} because it gives a measure of the effective particle-antiparticle asymmetry of a given species at any time. This should be compared to the primordial asymmetry $\eta$, which provides such information \emph{only} in the ultra-relativistic limit. The functional dependence of $r$ on the primordial asymmetry $\eta$ and the strength of the particle interactions, $\langle\sigma_{\textrm{ann}} v\rangle$, will be given below. The parameter $r$ is in principle an observable quantity as it controls the expected indirect signal coming from particle annihilation. Small $r$  corresponds to an extremely asymmetric case in which indirect annihilation signals are small or even absent, whereas in the large $r$ regime such signals are unsuppressed.

We assume that the Universe is radiation dominated in the epoch of interest, in which case the Hubble parameter is
\ba\label{H}
H(T)=\left(\frac{8 \pi^{3}}{90}\right)^{1/2} g_{\textrm{\small{eff}}}^{1/2}(T)\frac{T^2}{M_{Pl}}=\frac{1}{2t},
\ea
with $M_{Pl}\approx 1.22\times10^{19}$ GeV.

%The Jacobian of the transformation between $t$ and $x$ reads:
%\ba
%\frac{dt}{dx}=\left(\frac{90}{8 \pi^{3}}\right)^{1/2} \frac{M_{Pl}x}{ g_{\textrm{\small{eff}}}^{1/2}m^2}\left(1-\frac{1}{4}\frac{x}{g_\textrm{\small{eff}}}\frac{dg_\textrm{\small{eff}}}{dx}\right).
%\ea 
%{\bf Comment on $d g_eff/dx$}

If the annihilation cross section for the species $X$ is not too small compared to a typical weak scale process, the dynamical effects encoded in~(\ref{Boltz}) become relevant when the particle is non-relativistic. In this case the equilibrium distributions can be taken to be 
\ba
n_{eq}^\pm=n_{eq}e^{\pm\xi}\quad\quad\quad n_{eq}=g\left(\frac{mT}{2\pi}\right)^{3/2}e^{-m/T}.
\ea
In the above expression $g$ counts the internal degrees of freedom and $\mu=\xi T$ is the chemical potential. The explicit expression of the entropy density per comoving volume 
\ba
s(T)=\frac{2\pi^2}{45}h_{\textrm{\small{eff}}}(T)T^3,
\ea
allows us to introduce the convenient notation
\ba
Y_{eq}\equiv\frac{n_{eq}}{s}=ax^{3/2}e^{-x},~~~~~ x=\frac{m}{T},
\ea
with $a\equiv 45g/(4 \sqrt{2} \pi^{7/2}h_{\textrm{\small{eff}}}) $.

With these definitions the dynamical equation for $r(x)$ following from~(\ref{Boltz}) reads
\ba\label{Boltz'}
\frac{dr}{dx}&=&-\lambda\eta g_*^{1/2}\, x^{-n-2}\left[r-\frac{Y_{eq}^2}{\eta^2}(1-r)^2\right] \nonumber \\
&=& -\lambda\eta g_*^{1/2}\, x^{-n-2} \left[ r - r_{eq}\left( \frac{1-r}{1-r_{eq}}\right)^{2} \right], \ea
where 
\ba
\label{lambda}
\lambda&=&\left(\frac{\pi}{45}\right)^{1/2}M_{Pl}m\sigma_0,
\ea
and~\footnote{By requiring that $sR^3$ stays constant in time, taking into account the temperature dependence of $h_{\textrm{\small{eff}}}$ and using the relation $t=t(T)$ given by the definition of $H$, one finds that $d\log h_{\textrm{\small{eff}}}=\frac{3}{4}d\log g_\textrm{\small{eff}}$. With the help of this result one verifies that~(\ref{gstar}) coincides with the expression used in~\cite{darksusy}.} 
\ba
\label{gstar}
g_*^{1/2}=\frac{h_\textrm{\small{eff}}}{g_\textrm{\small{eff}}^{1/2}}\left(1-\frac{1}{4}\frac{x}{g_\textrm{\small{eff}}}\frac{dg_\textrm{\small{eff}}}{dx}\right).
\ea
In the above we have also defined $r_{eq} \equiv e^{-2\xi(x)}$, where $\xi$ is determined by 
\ba\label{xi} 
2\sinh\xi=\frac{\eta}{Y_{eq}}.
\ea
%For future reference a convenient expression for the equilibrium solution, obtained by inspecting (\ref{xi}), is 
%\ba\label{fo1}
%r_{eq}= \frac{Y_{eq}^2}{\eta^2}(1-r_{eq})^2,
%\ea

Notice from (\ref{gstar}) that we have taken into account the temperature dependence in $h_\textrm{\small{eff}}$ and $g_\textrm{\small{eff}}$. Because $g_\textrm{\small{eff}}$ is monotonically increasing with $T$, we see that the parentheses in the definition of $g^{1/2}_*$  is positive definite. In the numerical results that follow we use the data table from DarkSUSY~\cite{darksusy} for the temperature dependence of $g_{*}(T)$ and $h_{\textrm{\small{eff}}}(T)$.~\footnote{Note that we use the notation of DarkSUSY for the massless degrees of freedom parameters. To translate our notation to that of Kolb and Turner~\cite{KT} one should make the substitutions $g_\textrm{\small{eff}} \rightarrow g_{*}$ and $h_\textrm{\small{eff}} \rightarrow g_{*S}$.}

\begin{figure}%[t] %  figure placement: here, top, bottom, or page
\begin{center}
\includegraphics[width=4.5in]{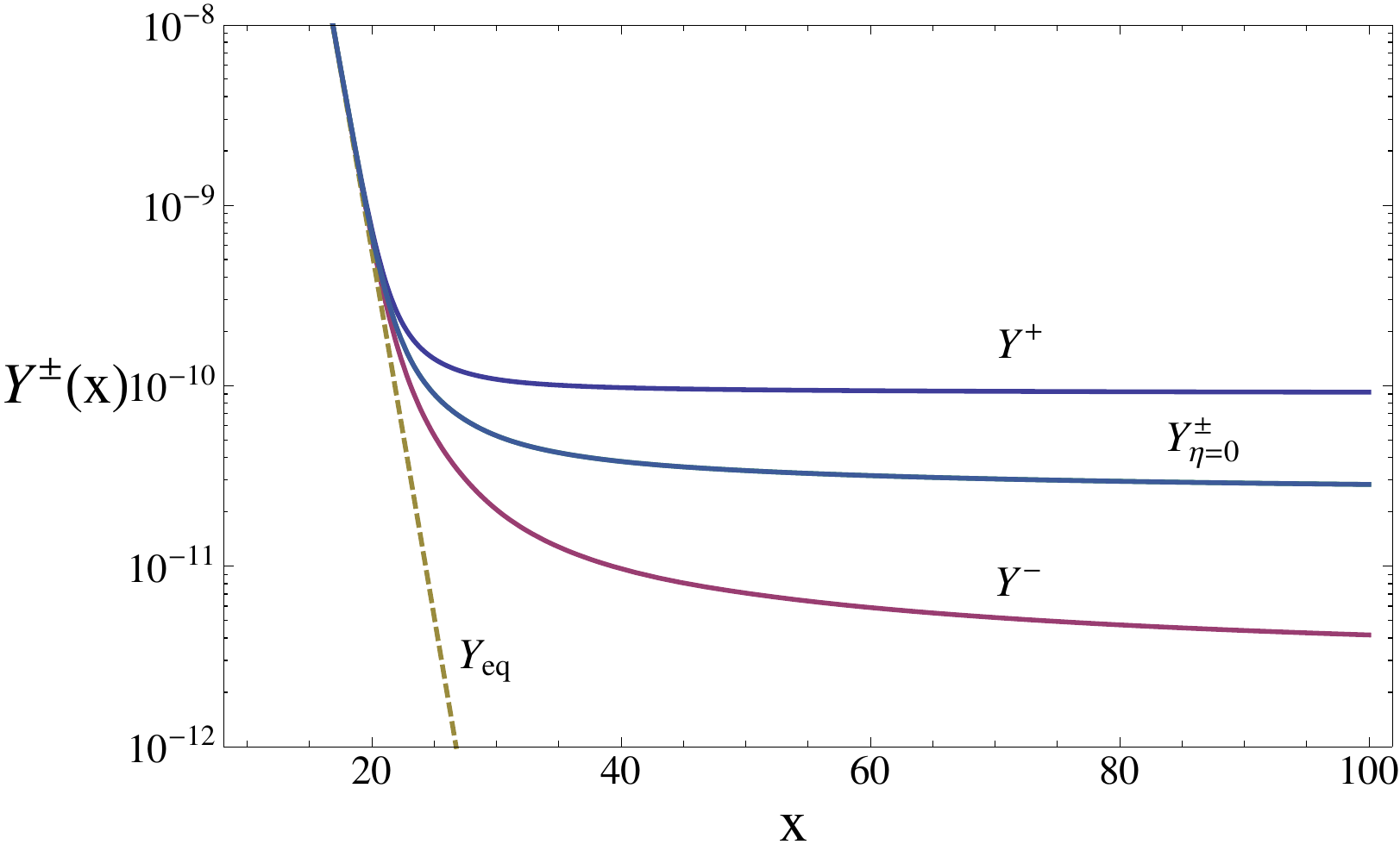}
\caption{\small Evolution of  $Y^{\pm}(x)$ illustrating the effect of the asymmetry $\eta$. After freeze-out both $Y^-$ and $Y^+$ continue to evolve as the anti-particles find the particles and annihilate. The $Y^\pm_{\eta =0}$ curve shows the abundance for $\eta=0$, a mass $m=10$ GeV and annihilation cross-section $\sigma_0=2$ pb. In contrast, with a non-zero asymmetry $\eta=\eta_B = 0.88 \times 10^{-10}$ and same mass and cross-section, the more abundant species (here $Y^+$) is depleted less than when $\eta=0$.  Also shown is the equilibrium solution $Y_{eq}(x)$.
\label{fig0}}
\end{center}
\end{figure}

Eq.~(\ref{Boltz'}) reproduces the well known case $\eta=0$ for which one finds that $r=1$ for any $x$. We will instead focus on scenarios with $\eta\neq0$ in the following. As shown in Fig.~\ref{fig0}, the effect of nonzero $\eta$ is to deplete the less abundant species more efficiently compared to $\eta = 0$ for the same annihilation cross section and mass.

%%%%%%%%%%%%%%%%%%%%%%%%%%%%%%%
%%%%%%%%%%%%%%%%%%%%%%%%%%%%%%

\subsection{The relic abundance of asymmetric species}
\label{analyticsol}

Equation~(\ref{Boltz'}) can be solved by numerical methods and imposing an appropriate initial condition at a scale $x=x_i\geq10$, where the non-relativistic approximation works very well. Although we have chosen $x_i=10$, we have checked that larger values ($10<x_i<x_f$, where $x_f$ is defined below) do not alter the final result. From (\ref{Boltz'}) one sees that in the early Universe $r = r_{eq}$, provided the cross-section is not too small. In our numerical solutions $r(x_i)$ is chosen to equal its equilibrium value $r(x_i)\equiv r_{eq}(x_i)=e^{-2\xi(x_i)}$.

We now present a very accurate analytic approximation of the numerical results. The numerical, exact solution will be compared to it shortly.

While there are no analytic solutions of~(\ref{Boltz'}),  it is not difficult to guess the qualitative behavior of $r(x)$. As already mentioned, for small $x$ the solution is very close to the equilibrium solution $r_{eq}(x)$. %, although $r(x)=r_{eq}(x)$ is not an exact solution for the obvious reason that $d r_{eq}/dx$ is non-vanishing.
 As the Universe expands, $r$ eventually has difficulty tracking the equilibrium expression which is exponentially decreasing. The easiest way to see this is to inspect the first equation in~(\ref{Boltz'}) and observe that the second term on the right-side is proportional to $Y_{eq} \sim e^{-x}$. The deviation of $r$ from the equilibrium solution begins to grow. 

The \textit{freeze-out} scale $x\equiv x_f$ is then defined by the condition that the two terms on the right hand side of~(\ref{Boltz'}) no longer balance each other, and $d r /dx \approx d r_{eq}/dx $ becomes comparable to the terms on the right-side of the Boltzmann equation. 
The behavior for $x>x_f$ is then dominated by the first term on the right hand side of~(\ref{Boltz'}) whatever the asymmetry is because of the exponential dependance of $Y_{eq}$ on $x$. We can therefore write
\ba\label{rx}
r(x)=r_{eq,f}\, e^{-\lambda\eta\,\Phi(x,m)}
\label{Exponentialr}
\ea
with
\ba\label{Phi}
\Phi(x,m)&\equiv& \int_{x_f}^x dx'\,  x'^{-n-2}g_*^{1/2}\,\\\no
&=& \int_{m/x}^{m/x_f} \frac{dT }{m}\left(\frac{T}{m}\right)^ng_*^{1/2} \\\no
\ea
and  $r_{eq,f} \equiv r_{eq}(x_f) $.
The present value of the fractional asymmetry is thus controlled by the quantities $r_{eq,f}$, $x_f$, $m$, and ultimately by $\xi_{f}$. Importantly, observe from (\ref{Exponentialr}) that the fractional asymmetry is \emph{exponentially} sensitive to the annihilation cross-section.

It remains to find an estimate for the freeze-out scale. An analytic expression for $x_f$ is obtained by the condition that $d r /dx \approx d r_{eq}/dx$ is comparable to either the first or second term in eq.~(\ref{Boltz'}).
This leads to 
\ba 
\frac{d r_{eq}}{d x } \approx -\delta\, \lambda \eta g^{1/2}_* x^{-n-2} r_{eq}, 
\ea 
or equivalently using the approximate relation~(\ref{xi}), 
\ba\label{FO}
\left(1-\frac{3}{2x_f}\right)\tanh\xi_f=\left(1-\frac{3}{2x_f}\right)\frac{1-r_{eq,f}}{1+r_{eq,f}}\approx\delta\frac{\lambda\eta g_{*,f}^{1/2}}{2 x_f^{n+2}},
\ea
with $\delta$ a number to be determined by fitting the numerical solution. Because $x_f$ turns out to be $\gtrsim20$ for the scenarios of interest (see the end of this section), we can ignore the $3/2x_f$ factor in the above expression.

As long as $g_{*,f}^{1/2}\lambda\eta<2x_f^{n+2}$, the condition~(\ref{FO}) can be solved iteratively. After one iteration, and with the help of equation~(\ref{xi}), we find
\ba\label{xf}
x_f\approx \log(\delta g_{*,f}^{1/2}a_f\lambda)+\frac{1}{2}\log\frac{\log^3(\delta g_{*,f}^{1/2}a_f\lambda)}{\log^{2n+4}(\delta g_{*,f}^{1/2}a_f\lambda)-g_{*,f}\left(\delta \frac{\lambda\eta}{2}\right)^2}+\dots.
\ea
This expression is accurate at the percent level, so we will use it in our applications. Notice that the effect of a nonzero asymmetry $\eta$ on the freeze-out temperature is very mild (see for example Fig.~\ref{fig0}). In fact, for all $g_{*,f}^{1/2}\lambda\eta<2x_f^{n+2}$ the scale $x_f$ is very close to the result found in the symmetric limit, in which case $\delta$ is typically taken to be $\delta=n+1$~\cite{ST}. As already emphasized, however, the effect of $\eta$ on the anti-particle population is considerable.

The asymmetry can only be considered small if $g_{*,f}^{1/2}\lambda\eta\ll2x_f^{n+2}$, in which case one has $Y^+_\infty\simeq Y^-_\infty\gg\eta$, as anticipated in the introduction. More generally, in the ``weak" coupling regime $g_{*,f}^{1/2}\lambda\eta<2x_f^{n+2}$ we find that to a high accuracy $r_{eq,f}\simeq1$. The final estimate for the present fractional asymmetry in this limit thus reads
\ba\label{estimate1}
r_\infty= e^{-\lambda\eta\,\Phi}\quad\quad \left[\textrm{if}\quad\frac{\lambda\eta  g_{*,f}^{1/2}}{2  x_f^{n+2}}<1%,~~~\rm{``moderate ~asymmetry"}
\right],
\ea
where $\Phi$ is understood to be $\Phi=\Phi(\infty,m)$. This result agrees with~\cite{GS}.

While it is clear that Eq.~(\ref{estimate1}) describes the regime where $r_{\infty} \lesssim 1$, it is less clear how far this approximation extends to the ``strong" coupling regime in which $r_{\infty} \ll1$.  In order to appreciate the range of validity of~(\ref{estimate1}), we estimate the integral~(\ref{Phi}) assuming that $g_*^{1/2}$ is approximately constant in the relevant range of temperatures. With this approximation one finds that $\Phi\approx g_{*,f}^{1/2}/[(n+1)x_f^{n+1}]$, and~(\ref{estimate1}) becomes
\ba\label{gg}
r_\infty\simeq\exp\left(-\frac{\lambda\eta g_{*,f}^{1/2}}{(n+1)x_f^{n+1}}\right)=\exp\left(-\frac{\lambda\eta g_{*,f}^{1/2}}{2x_f^{n+2}}\,\frac{2x_f}{(n+1)}\right).
\ea 
To see how far the approximation extends, first note that for  $r_{\infty} \gtrsim O(10^{-4})$, the exponent in~(\ref{gg}) should be at most $O(10)$. Since for a weak-scale cross section $x_f\sim20$, the upper bound on the exponent translates into an upper bound of $\lambda\eta g_{*,f}^{1/2}/[2x_f^{n+2}]<O(0.2)$, in agreement with our assumption~(\ref{estimate1}). We thus conclude that~(\ref{estimate1}) not only incorporates the intermediate limit anticipated in~(\ref{intermediate}), but also the ``strong" limit in which $r_{\infty} \gtrsim 10^{-4}$.

As $2x_f^{n+2}\rightarrow \lambda\eta g_{*,f}^{1/2}$ the fractional asymmetry gets further suppressed since now $r_{eq,f}$ can be much smaller than 1. Because in this regime $r_\infty$ is essentially vanishing for all practical purposes, we will be mainly concerned with the regime~(\ref{estimate1}) in the following sections.

%In the second regime $\tanh \zeta_f \simeq O(1)$ and the asymmetry can be considered ``large". That is, $2x_f^{n+2}\approx g_{*,f}^{1/2}\lambda\eta$. In this case one finds $r_f\simeq Y_{eq,f}^2/\eta^2$ from~(\ref{fo1}) and the final estimate for the present fractional asymmetry reads
%\ba\label{estimate2}
%r_\infty\simeq \frac{Y_{eq,f}^2}{\eta^2}e^{-\lambda\eta\,\Phi}\quad\quad \left[\lambda\eta\approx\frac{2x_f^{n+2}}{g_{*,f}^{1/2}},~~~\rm{``large~asymmetry"}\right].
%\ea

%%%%%%%%%%%%%%%%%%%%%%%%%%%%%%%
%\begin{figure}%[t] %  figure placement: here, top, bottom, or page
%\begin{center}
%\includegraphics[width=4.5in]{rx.pdf}
%\caption{\small Numerical solution (solid red lines) and analytical approximation (\ref{estimate1}) (dashed blue lines) for $r(x)$ compared. For illustration, the parameters chosen here are: (top curve) $m_X = 10$ GeV, $\eta =\eta_B$, $\sigma_0 = 5$ pb; (bottom curve) $m_X = 10$ GeV, $\eta =\eta_B$, $\sigma_0 = 7$ pb. Also note the exponential sensitivity of $r$ to the annihilation cross-section: here a modest change in $\sigma_0$ results in a factor of 10 change in $r_{\infty}$. WHY IS THE SUPPRESSION SO SMALL DESPITE THE LARGE CROSS SECTION?
%\label{fig0b}}
%\end{center}
%\end{figure}
%%%%%%%%%%%%%%%%%%%%%%%%%%%%%%

Finally, the estimate~(\ref{estimate1}) turns out to be a very good approximation of the exact (numerical) result. We scanned a large range of parameter space ranging from $\eta=O(10^{\pm4})\eta_B$, $m=O(10^{\pm4})m_p$, and $\sigma_0$ several orders of magnitude above and below the ``thermal WIMP" value $\sigma_{0,WIMP}=O(1)$ pb for both s-wave and p-wave processes.  Based on this analysis, we have found that our analytic expression departs by at most $5\%$ from the numerical result. Consistent with the derivation presented above, within the scanned range we have $\lambda\eta g_{*,f}^{1/2}/[2x_f^{n+2}]<O(0.1)$.

\section{Asymmetric WIMP dark matter}
\label{adm}

%In this scenario one recasts much of the formalism of baryogenesis in a dark sector such that one ends up with many more $X$ particles than $\bar{X}$ particles. This is in stark contrast with the thermal WIMP miracle which predicts a symmetric DM population $n_{X} = n_{\bar{X}}$. Note that ADM is doomed to never have any indirect signal coming from its present day annihilation signal since there are no antiparticles with which $X$ may annihilate. 

In asymmetric dark matter (ADM) scenarios the dark matter is assumed to have a primordial asymmetry $\eta$. For definiteness we will measure the DM asymmetry in units of the baryon asymmetry via the relation
\ba\label{eta}
\eta=\epsilon\eta_B
\ea
where $\eta_B = (0.88 \pm 0.021)\times10^{-10}$~\cite{WMAPetab} \footnote{$\eta$ and $\eta_B$ are normalized relative to the total entropy density (i.e., neutrinos $+$ photons).}.  The present-day abundance of baryons is $\rho_{B} = m_{p}~s~\eta_{B}$, whereas the DM abundance is
\ba \label{presconst} \rho_{DM} &=& m~s \left(Y^{+} + Y^{-} \right) \nonumber \\
 &=& m ~s \left( Y^{+} - Y^{-} + 2 Y^{-} \right) \no \\
 &=& m~s \left( \eta + 2 \frac{\eta ~r_{\infty} }{1- r_{\infty}}\right), \ea
where we have used~(\ref{YY}) to express the present day density of anti-DM, $Y^{-}$, in terms of $\eta$ and $r_\infty$.  The first term of (\ref{presconst}) corresponds to the asymmetric component which survives in the limit $r_{\infty} \rightarrow 0$, where the abundance scales as $\rho_{DM} \sim  \eta $.  Whereas the second term corresponds to the symmetric component, in which case the abundance scales as $\rho_{DM} \sim 1/ \langle \sigma_{ann} v \rangle$ when $r_{\infty} \rightarrow 1$.  In the intermediate regime, one may expect the total abundance to scale as simply the sum of these two approximate scalings.  However, this is not the case in general since the symmetric component depends exponentially on $\eta \, \langle \sigma_{ann} v \rangle$.

We write the observational constraint~(\ref{Omegas}) as
\ba\label{const}
\frac{m}{m_p}\epsilon = \left(\frac{1-r_{\infty}}{1+r_{\infty}}\right)\frac{\Omega_{DM}}{\Omega_B},
\ea
with $m_p$ and $m$ the proton and DM masses, respectively, and $r_\infty\equiv Y^-_\infty/Y^+_\infty$ the present fractional asymmetry.  The present-day constraints (\ref{presconst}) and $\rho_{B} = m_{p}~s~\eta_{B}$ are equivalent to satisfying (\ref{const}), generating the correct baryon asymmetry $\eta_B$, and having a theory for (\ref{eta}).

The literature so far has focused on the following two extreme scenarios. In the first scenario (ADM) the DM is totally asymmetric, namely $r_\infty=0$. The observed relic abundance is then explained if the DM mass is \emph{tuned} to the value
\ba\label{ADMm}
\textrm{ADM:}\quad m_{ADM}= \frac{\Omega_{DM}}{\Omega_B}\frac{m_p}{\epsilon},
\ea
and there is no strong constraint on the cross section. The cross section simply needs to be large enough (see below) to suppress $r_\infty$.

In the second extreme scenario (thermal WIMP) the DM is totally symmetric, namely $r_\infty=1$. From our analytic expression~(\ref{estimate1}) we find that in this limit the relation~(\ref{const}) becomes  
\ba\label{WIMP}
\textrm {Thermal WIMP:}\quad\quad\sigma_{0,WIMP}=\frac{1}{\eta_Bm_p}\frac{\Omega_B}{\Omega_{DM}}\frac{1}{M_{Pl}\Phi_{WIMP}}\sqrt{\frac{180}{\pi}}.
\ea
If $g_*$ is assumed to be constant, $\Phi=g_{*,f}^{1/2}/[(n+1)x_f^{n+1}]$, and the above formula becomes the one commonly used in the literature. The DM abundance is now determined by \emph{tuning} the annihilation cross section, the dependence on $m$ being very mild.

\begin{figure} %  figure placement: here, top, bottom, or page
\begin{center}
\includegraphics[width=3.5in]{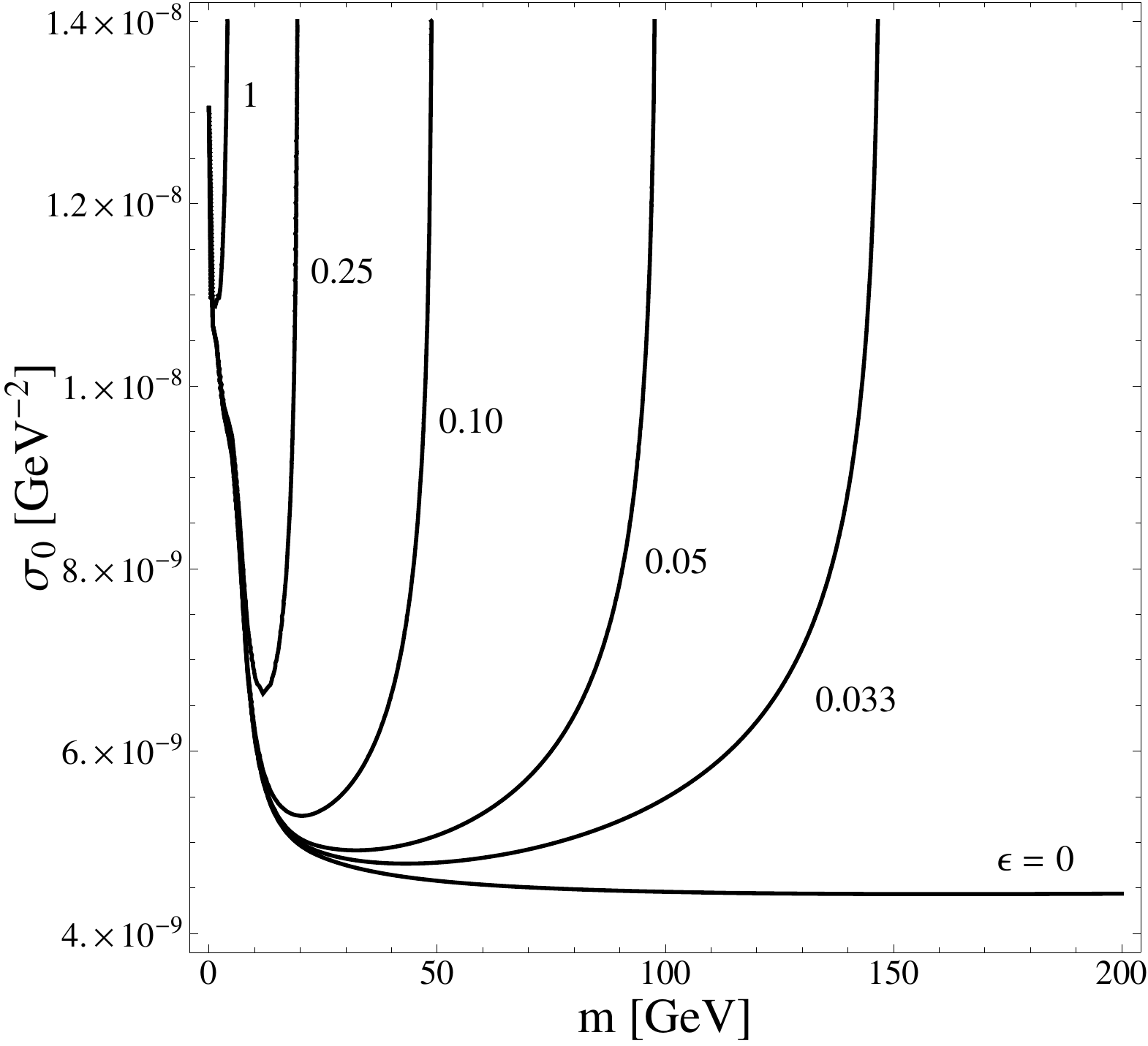}\\
\vspace{5mm}
\includegraphics[width=3.5in]{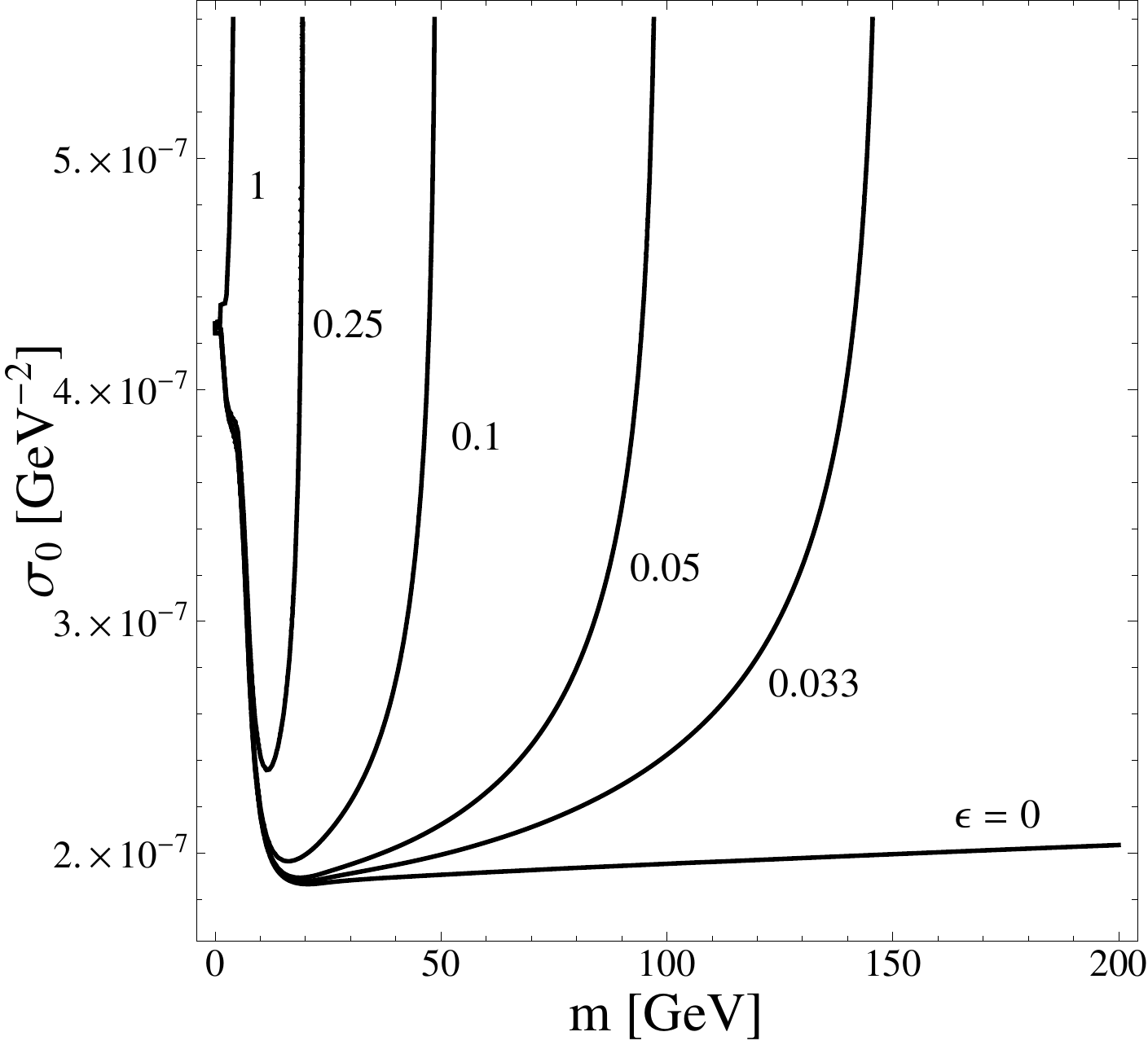}
\caption{\small Here we plot the annihilation cross section $\sigma_0$ required to reproduce the correct DM abundance $\Omega_{DM}$ via a s-wave process $n=0$ (above plot) and p-wave $n=1$ (bottom plot)  for a given dark matter mass $m$, and for various values of the primordial asymmetry $\eta=\epsilon\eta_B$. The line for $\epsilon=0$ corresponds to the usual thermal WIMP scenario. Notice that the fractional asymmetry runs from $r_\infty=0$ in the upper part of the curves to $r_\infty=1$ when the lines converge on the standard thermal WIMP curve. The effect of the QCD phase transition appears as a bump at $m\lesssim20$ GeV, as anticipated in the text.  Note that the bottom plot is basically enhanced by a factor $\Phi_{n=0}/\Phi_{n=1}\sim (n+1)x_f$ compared to the former.  As a reference, recall that $1$ pb $\simeq2.6\times10^{-9}$ GeV$^{-2}$.
\label{fig1}}
\end{center}
\end{figure}

For the intermediate scenarios in which $0<r_\infty<1$ the physics interpolates between the above extremes. In Fig.~\ref{fig1} we plot the allowed parameter space in the $m-\sigma_0$ plane of a DM candidate satisfying the condition~(\ref{const}) for different choices of the asymmetry parameter $\epsilon$ (see Eq.~(\ref{eta})). The two figures refer to purely s-wave and p-wave processes, respectively. For a given asymmetry $\epsilon$, we see that the allowed region is bounded from below by the WIMP cross section $\sigma_{0,WIMP}$~(\ref{WIMP}), and bounded from the right as a maximum on the mass at the ADM mass value~(\ref{ADMm}).

For $\epsilon\rightarrow0$ and $m\gtrsim20$ GeV the lines asymptote to a horizontal line representing the thermal WIMP scenario, where the cross section is fixed but the mass can vary. The regime $m\lesssim20$ GeV is instead strongly sensitive to the abrupt change in the number of degrees of freedom around the QCD phase transition. The latter appears as a large gradient in $g_*^{1/2}$ at $T^*\lesssim1$ GeV~\cite{darksusy}. The effect on the relic abundance of the species is relevant if $T_f=m/x_f\lesssim T^*$, and can be seen in Fig.~\ref{fig1} as the bump at $m\lesssim x_fT^*\simeq20$ GeV. This change in the degrees of freedom also tends to reduce the exponent in~(\ref{rx}) for light species.

\begin{figure}%[t] %  figure placement: here, top, bottom, or page
\begin{center}
\includegraphics[width=4.5in]{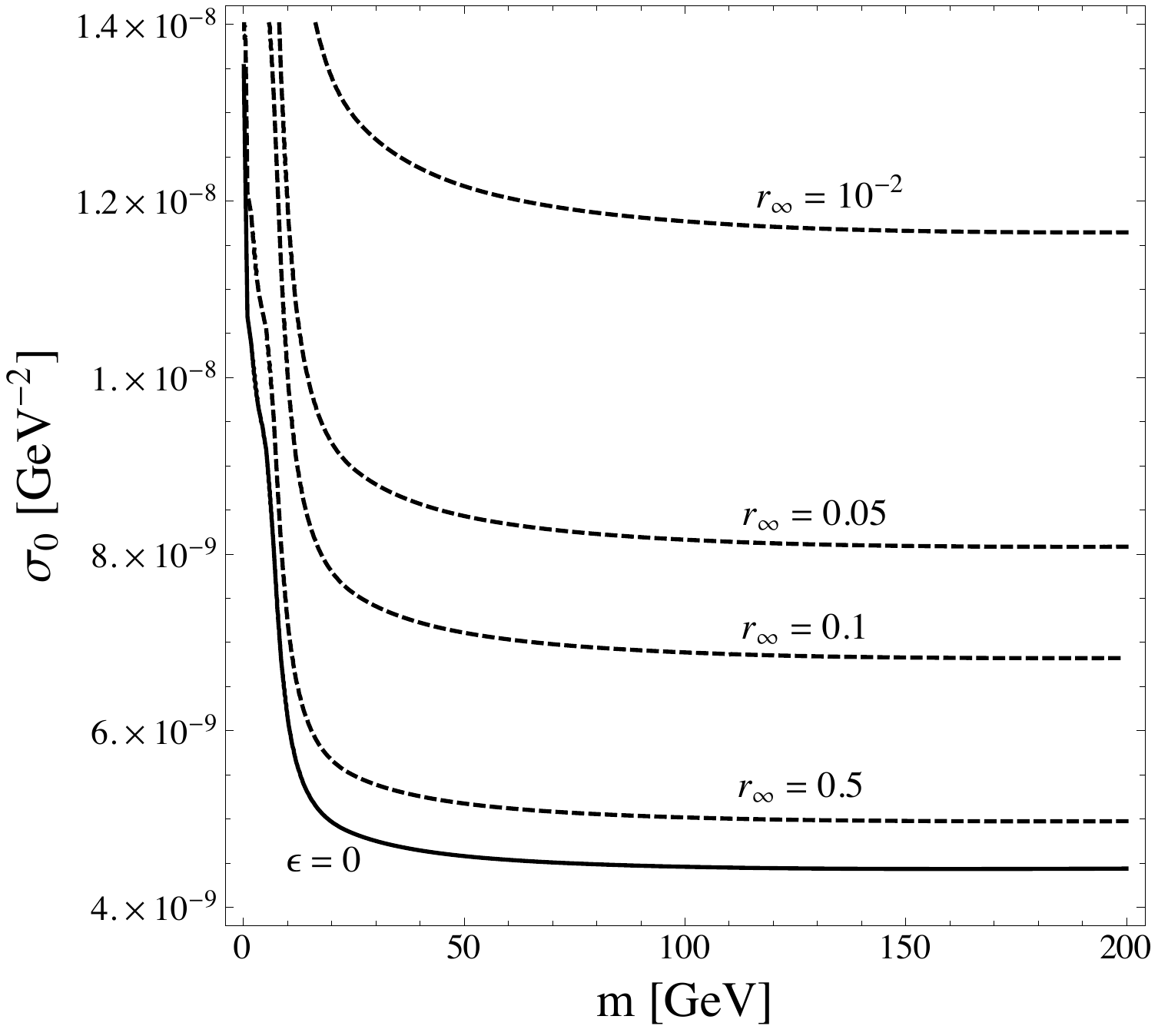}
\caption{\small Same as in Figure \ref{fig1} showing contours of constant $r_{\infty}$, for s--wave process. The completely asymmetric scenario $r_{\infty} \ll 1$ corresponds to the top region of the plot having cross-sections always larger than the thermal WIMP cross-section. For reference we show the $\epsilon=0$ line which corresponds to the usual thermal WIMP scenario ($r_{\infty}=0$).  \label{fig2b}}
\end{center}
\end{figure}

In Fig.~\ref{fig1} the fractional asymmetry varies from $r_\infty=0$ in the upper part of the curves to $r_\infty=1$ when the curves overlap with the thermal WIMP scenario ($\epsilon=0$). The same information is presented differently in Fig. \ref{fig2b}, where we trade $\epsilon$ for $r_{\infty}$. % In Fig. \ref{fig2b} we see that values of $r_\infty$ smaller than $O(10^{-2})$ correspond to annihilation cross-sections only a factor of a few larger than the typical thermal WIMP cross-section needed to obtain the observed dark matter abundance for a symmetric species [for more details, see below]. 
In this figure the exponential sensitivity of $r_{\infty}$ to the annihilation cross-section is evident.

%Interestingly, notice that for a given cross section there are typically two DM masses that account for the observed relic abundance whenever $\epsilon\neq0$.

%%%%%%%%%%%%%%%%%%%%%%%%%%%%%%%
%%%%%%%%%%%%%%%%%%%%%%%%%%%%%%

%%%%%%%%%%%%%%%%%%%%%%%%%%%%%%%
%\begin{figure} %  figure placement: here, top, bottom, or page
%\begin{center}
%\includegraphics[width=3.5in]{SigmaMn1.pdf}
%\caption{\small Same as in Fig.\ref{fig1} but for a p-wave process ($n=1$). Note that the latter plot is basically enhanced by a factor $\Phi_{n=0}/\Phi_{n=1}\sim (n+1)x_f$ compared to the former. \label{fig2}}
%\end{center}
%\end{figure}
%%%%%%%%%%%%%%%%%%%%%%%%%%%%%%

A generic model for beyond the SM physics is expected to lie in between the two extreme scenarios described above. In fact as emphasized in the introduction, a primordial DM asymmetry ($\epsilon\neq0$) is generally expected; meanwhile most models beyond the SM naturally lead to WIMP-like cross sections. The resulting \emph{asymmetric WIMP} scenario will typically have $\sigma_0\gtrsim\sigma_{0,WIMP}$ and $r_\infty\lesssim1$. %As can be seen in Figs.~\ref{fig1},~\ref{fig2}, and \ref{fig2b}, [???] the region of parameter space characterizing this class of models becomes larger and larger as the asymmetry decreases [this will become more evident when discussing explicit models; see next subsections].  
%For a given $\epsilon$ the allowed region is bounded from below by the WIMP %plateau at $\sigma_{0} \sim 4\times10^{-9}$ GeV$^{-2}$ for $n=0$ and $\sim2\times10^{-8}$ GeV$^{-2}$ for $n=1$, 
%value $\sigma_{0,WIMP}$~(\ref{WIMP}), and bounded from the right as a maximum on the mass at the ADM mass value~(\ref{ADMm}). %Notice also that the thermally averaged cross section for an asymmetric DM candidate is always bigger than that of an exactly symmetric candidate of the same mass. 

It is instructive to re-express the fractional asymmetry for an asymmetric DM species as a function of its thermally averaged cross section $\sigma_0$ and the quantity $\Phi=\Phi(\infty,m)$ defined in~(\ref{Phi}): 
\ba\label{rinf}
r_{\infty}= e^{-\lambda\eta\Phi}=\exp\left[-2\left(\frac{\sigma_0}{\sigma_{0,WIMP}}\right) \left(\frac{\Phi}{\Phi_{WIMP}}\right) \frac{1-r_{\infty}}{1+r_{\infty}}\right],
\ea
where $\sigma_{0,WIMP}$ and $\Phi_{WIMP}\approx\Phi$ refer to a symmetric DM candidate of the same mass. The first equality in~(\ref{rinf}) is a good analytic approximation to the numerical solution (see~(\ref{estimate1})), whereas the second is exact and follows from eliminating $\eta$, $\lambda$ and $m$ using (\ref{lambda}) and the empirical formul\ae~(\ref{const}) and~(\ref{WIMP}). 

\begin{figure}%[t] %  figure placement: here, top, bottom, or page
\begin{center}
\includegraphics[width=4.5in]{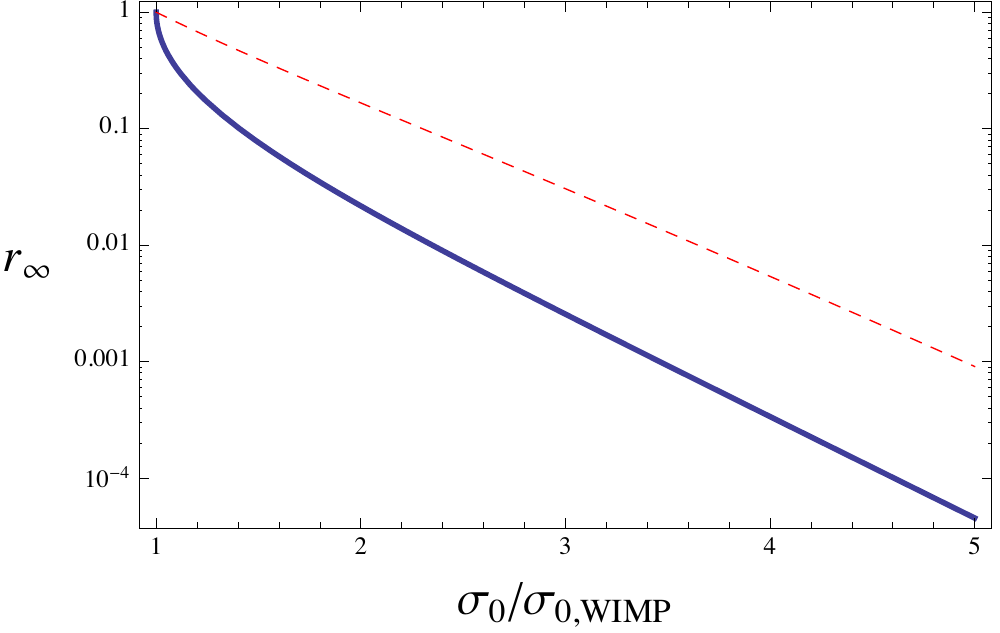}
\caption{\small Present fractional asymmetry $r_\infty=Y^-_\infty/Y^+_\infty$ (see (\ref{rinf})) for various values of the thermally averaged cross section $\sigma_0$ (defined as $\langle\sigma_{\textrm{\small{ann}}}v\rangle\equiv\sigma_0 (T/m)^n$ in the first section) in units of the value $\sigma_{0,WIMP}$ obtained for an exactly symmetric species of the same mass, see eq.(\ref{WIMP}). All the points in the curve, i.e. all solutions of~(\ref{rinf}), account for the observed DM density. The dashed line is the suppression factor~(\ref{sup}) for indirect detection. \label{fig3}}
\end{center}
\end{figure}

From (\ref{rinf}) we  see explicitly that for WIMP-like cross-sections and larger, the dependence of $r_{\infty}$ on the cross section is very sensitive. %The present-day value $r_\infty$ is quite sensitive to such values of the cross-section. 
For a more quantitative evaluation,  we show in Fig.~\ref{fig3} the solution of the above implicit equation. Indeed, we see that an $O(1)$ departure from the thermal WIMP cross section implies a significant change in $r_\infty$.

Indirect signals of annihilating asymmetric DM in the Universe are suppressed compared to those of a symmetric thermal candidate of the same mass by the quantity 
\ba\label{sup}
\frac{\sigma_0}{\sigma_{0,WIMP}}\times r_\infty\times \left(\frac{2}{1+r_\infty}\right)^2 \leq1.
\ea
One can understand the above formula as follows. The rate for indirect events is proportional to $\langle\sigma_{\textrm{\small{ann}}}v\rangle r_\infty(n^+_{\eta\neq0})^2$ for an asymmetric candidate, or similarly to $\langle\sigma_{\textrm{\small{ann}}}v\rangle (n^+_{\eta=0})^2$ for a symmetric (thermal WIMP) candidate. Notice, however, that the number densities $n^+_{\eta\neq0}$ and $n^+_{\eta=0}$ are not the same. In fact, for DM candidates of the same mass it is the total (particle plus antiparticle) densities that must be equal, namely $n_{\eta\neq0}^+(1+r_\infty)=2n^+_{\eta=0}$. Taking the ratio of these rates gives~(\ref{sup}). 

We plot the quantity~(\ref{sup}) as a function of the cross section as a dashed line in Fig.~\ref{fig3}. Interestingly, we find that potentially important \emph{indirect} signals are still effective in the intermediate regime $\sigma_0\gtrsim\sigma_{0,WIMP}$ in which $r_\infty\lesssim1$. For example, a suppression of $\,\lesssim0.5$ is found for values of the fractional asymmetry below $\sim0.1$ (i.e. for $\sigma_0/\sigma_{0,WIMP}\gtrsim1.5$), when the DM could already be considered asymmetric.

%%%%%%%%%%%%%%%%%%%%%%%%%%%%%%%
%%%%%%%%%%%%%%%%%%%%%%%%%%%%%%

%%%%%%%%%%%%%%%%%%%%%%%%%%%%%%%
%%%%%%%%%%%%%%%%%%%%%%%%%%%%%%

\section{Asymmetric WIMP dark matter: applications}\label{adm:app}

%From Figures \ref{fig1} and \ref{fig2} it is clear that the allowed region of cross-sections and masses is increased when there is a non-zero fractional asymmetry.

In the following subsections we apply our results to a few ADM models, choosing our examples from among those already existing in the literature~\cite{ADM,Xogenesis}. We demonstrate there is a large allowed region in these models in which the present-day fractional dark matter asymmetry is in the range $0<r_\infty<1$ and, {\em significantly}, the dark matter annihilation cross-section in the early Universe can be comparable to the typical WIMP cross-section. Our results generalize the conclusions found in the previous literature, which only discusses completely asymmetric dark matter, corresponding to a present fractional asymmetry of $r_\infty=0$ in our notation.  
%Because the "WIMP miracle" appears to be more effective for DM masses around the weak scale we focus on scenarios in which the Boltzmann suppression in the transfer mechanism  (see~(\ref{boltz}) below) is active~\cite{Xogenesis}. 
%Compared to~\cite{Xogenesis}, the discussion presented here generalize the results to $r_{\infty} >0$, and include the effect of solving the Boltzmann equation, which is necessary to relate the present-day abundances of dark matter and baryons when $r_{\infty } \neq 0$ (see (\ref{const})). 

In the examples we discuss the MSSM is assumed to couple to dark matter superfields $X,\bar X$ via a heavy messenger sector. The dark matter chiral superfields $X, \overline{X}$ carry $\pm 1$ charge under a hidden sector global $U(1)$, are Standard Model gauge singlets, and have a SUSY invariant mass $m_X$. For definiteness we assume that after SUSY breaking the fermionic component $\psi_X$ is the dark matter. Similar results hold for scalar DM candidates. 

The overall system is assumed to violate either lepton ($L$) number or baryon ($B$) number together with the dark ($X$) number, but to (perturbatively) preserve a linear combination of these symmetries. When the number-violating interactions are in thermal equilibrium, chemical equilibrium forces a relation between the asymmetries of the visible and dark sectors and therefore determine the quantity $\epsilon$ introduced earlier~(\ref{eta}). We will refer to this effect as a transfer mechanism.

\subsection{Example 1: Lepton-number violating transfer operator}
\label{example1}

In the first example we consider lepton number violation and take the leading transfer mechanism between the dark and leptonic sectors, after having integrated out the messengers, to be described by a nonrenormalizable term in the superpotential~\cite{ADM}:
\ba\label{transf3}
\Delta W_{asym}= \frac{XXH_uL}{\Lambda},
\ea
with $\Lambda$ taken to be close to the TeV scale. The operator~(\ref{transf3}) preserves a global $L'=X-2L$ number, but violates both $L$ and $X$. The quantum numbers $L,X$ become good quantum numbers in the early Universe as soon as the transfer operator~(\ref{transf3}) freezes out, assumed to occur for simplicity at an instantaneous temperature $T_D$. One can show that for $\Lambda=O(1)$ TeV $T_D$ is below the scale $T_{sph}$ at which the sphalerons shut off, in which case the only source of lepton number violation at low temperatures is induced by~(\ref{transf3}). 

This model has either one or two stable particles, depending on the relative mass difference between the $\psi_X$'s and the lightest superpartner of the SM. To see that note that the model has a $Z_4$ symmetry (see also~\cite{ADM}) which is the usual $Z_2$  $R-$parity in the visible sector, plus a $Z_4$ in the $X$ sector that acts as $+i$ on the fermionic component $\psi_X$ and $-i$ on the scalar component $\widetilde{X}$ (the $\overline{X}$ fields have the opposite charge). This symmetry remains unbroken by soft SUSY and electroweak symmetry breaking, and guarantees that at least one component of $X$ is always stable. The other SUSY component of $X$ generically decays through the transfer operator to the lightest component of $X$ and SM fields.  

The transfer operator typically causes the lightest superpartner (LSP) of the SM to be unstable, if the channel is kinematically allowed. (The decay $\hbox{LSP} \rightarrow  XX + SM ~particles$ is consistent with the $Z_4$ symmetry).
If $m_{LSP} < 2 m_X$ then this decay is forbidden and the model has two stable particles.  In what follows we will assume $m_{LSP} > 2 m_X$ for simplicity.

%At these temperatures the only particles present in the thermal bath are the dark matter and Standard Model particles (minus the top quark).  

\subsubsection{Chemical Potential Analysis} 

To estimate the primordial DM asymmetry parameter $\epsilon$, see~(\ref{eta}), we proceed in two steps.  For a useful review on chemical potential analysis see~\cite{HT}. 

As a first step we impose chemical equilibrium and charge neutrality at the scale $T=T_{sph}>T_D$ at which the sphalerons decouple, assumed to be below the Higgs condensate scale $T_{c}$.  At these temperatures the only dynamical particles are the DM and the Standard Model fields (including the top quark), and we have:
\ba\label{chem}
Q&\propto&18\mu_u-12\mu_d-6\mu_e=0\\\no
\mu_u-\mu_d&=&\mu_\nu-\mu_e\\\no
\mu_u+2\mu_d+\mu_\nu&=&0\\\no
2\mu_X+\mu_\nu&=&0,
\ea
where the chemical potentials $\mu_u, \mu_d, \mu_\nu, \mu_e,$ and $\mu_X$ refer to the Standard Model up, down quarks, the (purely left handed) neutrinos, the charged leptons, and the DM respectively. The first equation in~(\ref{chem}) imposes electrical neutrality, the second accounts for the $W^\pm$ exchange, the third for the sphaleron process, and finally the last equation follows from the transfer operator~(\ref{transf3}). The solution of the system~(\ref{chem}) can be used to derive the primordial baryon and $L'$ asymmetries:
\ba\label{BL'}
\eta_B&= &-\frac{36}{7}\mu_e\\\no
\eta_{L'}&=&\left(\frac{25}{6}+\frac{11}{36} \frac{f(m/T_{sph})}{f(0)}\right)\eta_B,
\ea
where the function $f(x)$ is defined by
\ba\label{boltz}
f(x)=\frac{1}{4\pi^2}\int_0^\infty dy \frac{y^2}{\cosh^2\left(\frac{1}{2}\sqrt{x^2+y^2}\right)},
\ea
and accounts for the possibility that the DM be non-relativistic at the relevant temperature~\cite{Barr:1990ca}. If the DM is instead bosonic the $\cosh(x)$ function should be replaced by $\sinh(x)$. %The Boltzmann suppression in~(\ref{boltz}) is exponential if $x\geq5$, of the order of a ten percent if $1<x\lesssim5$, and absent for lighter masses. 

Below the sphaleron temperature the asymmetries for $B$ and $L'$ given above are conserved, while $L$ and $X$ are still violated by~(\ref{transf3}). In the second step of our chemical potential analysis we thus compute the present asymmetry for the $X$ number at the temperature $T_D$ at which the transfer operator decouples.

Chemical equilibrium at $T_D$ imposes a number of conditions similar to~(\ref{chem}). The two main differences are that at this scale the sphaleron process (see third line in~(\ref{chem})) is no longer effective, and that the top quark is not included. The system of equations for the chemical potentials is now solved by replacing the sphaleron process constraint with the initial conditions provided by~(\ref{BL'}), and not including the top quark contribution.  The result is finally
\ba\label{epsilonHuLXX}
\epsilon\equiv\frac{\eta_X}{\eta_B}=\left(\frac{309+22 \frac{f(m/T_{sph})}{f(0)}}{1026+72 \frac{f(m/T_D)}{f(0)}}\right)\frac{f(m/T_D)}{f(0)}.
\ea
%In this second step we assumed that the top is decoupled [this assumption is justified a posteriori]. 
Our result~(\ref{epsilonHuLXX}) reduces to the one found in~\cite{ADM} if the top quark is instead decoupled above the sphaleron scale, and if one assumes $f(m/T_D)=f(m/T_{sph})=f(0)$. We also checked that the result changes only mildly if one assumes that the massive vector bosons are decoupled at the scale $T_D$. In all these cases one finds that~(\ref{epsilonHuLXX}) can be well approximated as $\epsilon\approx0.3 f(m/T_D)/f(0)$.

\subsubsection{Transfer Operator Decoupling}

In order for the analysis of the previous sections to be applicable, the freeze-out temperature $T_f$ must be below the scale $T_D$ at which the transfer operator decouples~\footnote{We will comment on the limiting case $T_D=T_f$ in Section~\ref{discuss}. For the moment, notice that the case $T_D<T_f$ is clearly meaningless: freeze-out always occurs when the strongest interaction decouples.}: only under this condition there exists a temperature range $T_f\leq T\leq T_D$ in which Eq.~(\ref{Boltz}), and the subsequent analysis, is fully reliable. We will see in subsection~\ref{anndd} that, in typical UV completions of the model~(\ref{transf3}), this condition is ensured by higher dimensional operators that preserve the dark number and $L$, and which are therefore not involved in transferring asymmetries.

%The physical picture is hence the following. In the early Universe the SM is in thermal and chemical equilibrium with the supersymmetric partners, the DM, and the messenger sector. At very high scales the heavy fields decay, and the dynamical fields left in the bath are the SM particles and the DM. Notice that under the assumption that $m_{LSP}>2 m_X$ the LSP is also unstable, and decays via the transfer operator~(\ref{transf3}). In the specific case in which the neutralino is the LSP, the dominant decay modes are $\tilde{\chi} \rightarrow\overline{\psi}_X \overline{\psi}_X \overline{\nu}$ and the conjugate $\tilde{\chi} \rightarrow{\psi}_X {\psi}_X{\nu}$. 

The Boltzmann equation governing the evolution of the $X$ number at $T \gtrsim T_{D}$ is given by Eq.~(\ref{Boltz}), but with additional, $X$-number changing terms induced by~(\ref{transf3}). These latter involve the DM and the SM particles as final states, with typical reactions of the form 
\ba\label{XXnu'}
\psi_X \psi _X \leftrightarrow \overline{\psi}_X \overline{\psi}_X \overline{\nu} \overline{ \nu}.
\ea
The decoupling temperature $T_D$ is then estimated as the scale at which the strongest interaction mediated by the transfer operator -- formally described by a ``collision term" $C$ on the right hand side of~(\ref{Boltz}) -- becomes comparable to the Hubble expansion rate, namely when
\ba\label{TD}
C=\,H(T_D)n_X,
\ea
with $n_X$ the DM equilibrium number density.

If the transfer operator decouples when the DM is relativistic, the Boltzmann suppression in~(\ref{epsilonHuLXX}) is not effective and the parameter $\epsilon$ does not depend on the DM mass. In this case the details of the transfer mechanism are irrelevant. %For completeness we mention that in this limit the process~(\ref{XXnu'}) occurs via the exchange of a sneutrino and a neutralino~\cite{ADM}, with the latter particles being produced typically off-shell.

The Boltzmann suppression in~(\ref{epsilonHuLXX}) is effective if the transfer operator decouples when the DM is non-relativistic. In this case only the precise value of $T_D$, and hence the details of the transfer operator matter. In the specific case in which the neutralino is the LSP, we find that the dominant number-violating interaction mediated by~(\ref{transf3}) in the non-relativistic limit is associated to the process~(\ref{XXnu'}), and occurs via the production of an on-shell intermediate neutralino which eventually decays with a branching ratio BR $=1/2$ into an $X$- and $L$-number violating final state $\psi_X \psi _X \rightarrow \tilde{\chi} \overline{\nu}\rightarrow\overline{\psi}_X \overline{\psi}_X \overline{\nu} \overline{ \nu}$. %The reader is referred to Appendix~\ref{lhu} for more details. The collision term for the latter process can be written as
%\ba
%C&=&\frac{1}{2}\times\frac{1}{2}\int\frac{d^3p_{X_1}}{(2\pi)^32E_{X_1}}\int\frac{d^3p_{X_2}}{(2\pi)^32E_{X_2}}\int\frac{d^3p_{\tilde{\chi}}}{(2\pi)^32E_{\tilde{\chi}}}\int\frac{d^3p_{\overline{\nu}}}{(2\pi)^32E_{\overline{\nu}}}\\\no
%&\times&(2\pi)^4\delta^{(4)}(p_{X_1}+p_{X_2}-p_{\tilde{\chi}}-p_{\overline{\nu}})\,\langle | {\cal M}_{\psi_X \psi_X \widetilde{\chi} \nu} |^2 \rangle\,\,f_{X_1}f_{X_2},
%\ea
%where $\langle | {\cal M}_{\psi_X \psi_X \widetilde{\chi} \nu} |^2 \rangle$ denotes the spin averaged squared matrix element for the production $\psi_X \psi _X \rightarrow \tilde{\chi} \overline{\nu}$, whereas $f_i=e^{-E_i/T}$ is the thermal distribution for a particle species $i$ of energy $E_i$ at the temperature $T$. 

%[the error in the final formula implied by this approximation is estimated to be $O(m_X^2/m_{\tilde{\chi}}^2)<0.25$ and is therefore acceptable], %
If we neglect the DM mass, then at leading order in $T/m_{ \tilde{\chi}}$ the collision term $C$ reduces to (see Appendix \ref{lhu})
\ba\label{C}
C=-\frac{\sqrt{2\pi}}{4(2\pi)^5}\left(\frac{g_w v_u}{m_{\tilde\nu}^2\Lambda}\right)^2\, m_{ \tilde{\chi}}^8\left(\frac{T}{m_{ \tilde{\chi}}}\right)^{9/2}\,e^{-m_{ \tilde{\chi}}/T}.
\ea
In the above expression, $g_w$ is a weak gauge coupling, $v_u=v\sin\beta$ is the vacuum expectation value of the up-type Higgs, and $m_{\tilde\nu}$ is the sneutrino mass. The exponential suppression in~(\ref{C}) reflects the fact that the process occurs far in the Boltzmann tail of the thermal distribution of the incoming DM particles. 

The factors of $T$ in (\ref{C}) arise from a partial integration in the process of taking the thermal average, and reflect the fact that the phase space integral vanishes when the energy of the DM particles is at threshold. Another way to understand these factors of $T$ is to note that for the inverse process $\nu \tilde{\chi} \rightarrow \psi_{X} \psi_{X}$ the powers of $T$ arise solely from the number densities of the incoming particles simply because there is no threshold and the matrix element is set by the mass of the LSP.  The inverse collision term is therefore proportional to $n_{\tilde{\chi}} n_{\nu}\sim T^{3/2} T^{3} e^{-m_{\tilde{\chi}}/T}$.

%Because the precise value of the decoupling temperature is relevant only when $T_D\lesssim m_X/5$, in which limit~(\ref{C}) gives the dominant contribution, we will estimate $T_D$ in this model  

The decoupling temperature is finally estimated using~(\ref{TD}), with $C$ given by~(\ref{C}) and $n_X$ the non-relativistic DM equilibrium density. We find that, up to negligible logarithmic corrections, $T_D$ is given by 
\ba
T_D\approx \frac{m_{ \tilde{\chi}}-m_X}{\log\gamma},~~\quad\gamma=\frac{1}{4(2\pi)^3}\left(\frac{g_wv_u}{m_{\tilde\nu}^2\Lambda}\right)^2\frac{(m_{ \tilde{\chi}}-m_X) M_{Pl}m_{ \tilde{\chi}}^{7/2}}{\left(\frac{8 \pi^{3}}{90}\right)^{1/2} g_{\textrm{\small{eff,D}}}^{1/2}\,m_X^{3/2}}
\ea 
which approximately reads $T_D\sim(m_{ \tilde{\chi}}-m_X)/20$ if the mass scales are of $O($GeV-TeV$)$. We now have all the ingredients required to estimate the primordial asymmetry parameter $\epsilon$, see eq.~(\ref{epsilonHuLXX}), as a function of the DM mass and the LSP mass $m_{ \tilde{\chi}}$. This estimate has been found here working under the assumption that the transfer operator decouples when the DM is already non-relativistic. Yet, because the Boltzmann suppression in~(\ref{epsilonHuLXX}) ceases to be effective when the non-relativistic limit is still a good approximation, it is clear that our estimate of $\epsilon$ should also apply to the case in which the transfer operator decouples when the DM is relativistic. As emphasized above, indeed, in this latter case the details of the transfer mechanism do not matter.

\subsubsection{\label{discuss}Discussion} 

The main purpose of the present subsection is to scan the parameter space of the model~(\ref{transf3}) and show that there is a sizeable region in the allowed parameter space with a present-day fractional asymmetry in the range $0.1<r_\infty<1$. This is the asymmetric WIMP regime. 

%In this regime one can find $r_\infty\sim0.5$, in which case indirect signals are unsuppressed, and a very tiny fractional asymmetries can be naturally obtained with cross sections of weak-scale magnitude, without the need of including additional ingredients. %In order to illustrate these points we focus on the regime $m>T_D$, where one finds that small variations of the fundamental parameters result in....

\begin{figure}%[t] %  figure placement: here, top, bottom, or page
\begin{center}
\includegraphics[width=4.5in]{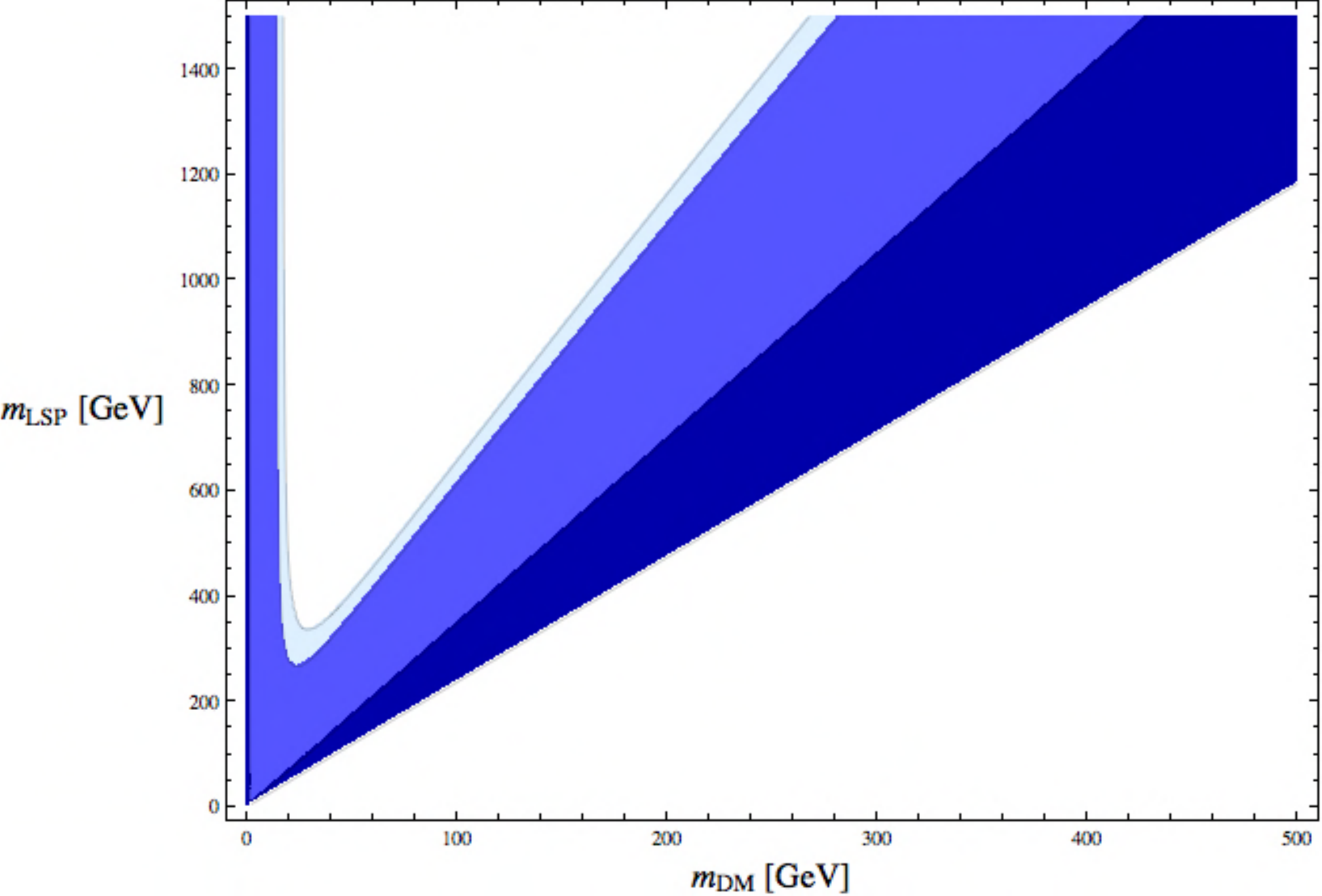}
\caption{\small Contour plot for $r_\infty$ in the allowed parameter space for the model~(\ref{transf3}). Below the lower line and above the upper line the DM density is smaller and higher than experimentally observed, respectively. In the dark blue area $0.9\leq r_\infty<1$, in the blue area $0.1\leq r_\infty\leq0.9$, and in the light blue area $0\leq r_\infty\leq0.1$. See the text for more details.  Here we have taken $m_{\tilde{\nu}}  = \Lambda = 1$ TeV and $\tan \beta = 20$.   \label{figHuLXX}}
\end{center}
\end{figure}

The relevant parameters of the model~(\ref{transf3}) at low temperatures reduce to the decoupling temperature $T_D$, the DM mass $m_X$, and the annihilation cross section (encoded in $r_\infty$). The requirement that the particle $\psi_X$ saturates the observed DM density is implemented by~(\ref{const}) and it amounts to a constraint on these parameters. In Fig.~\ref{figHuLXX} we show the resulting contour plot for the quantity $r_\infty$ \emph{defined} by~(\ref{const}) as a function of $m_X$ and $m_{ \tilde{\chi}}$.

%%%%%%%%%%%%%%%%%%%%%%%%%%%%%%%
%%%%%%%%%%%%%%%%%%%%%%%%%%%%%%

The upper curve shows the line~(\ref{ADMm}) at which $r_\infty=0$. In the area above this curve the DM abundance exceeds the observed value. This region of the parameter space is therefore excluded. The lowest curve is determined by the condition $T_D= T_f\sim m_X/20$, which approximately reads $m_{ \tilde{\chi}}\sim2 m_X$ in our case, at which $r_\infty\simeq1$ (and $\epsilon\ll1$). In the area below this latter curve the relic abundance of the species $\psi_X$ is smaller than the observed value. This area is excluded here because $\psi_X$ is assumed for simplicity to be the only DM candidate. The intermediate lines in Fig.~\ref{figHuLXX} separate regions with different present-day fractional asymmetry $r_\infty$.

The plot can be understood as follows. For any $r_\infty$, a given value for the DM mass completely determines $\epsilon$ with~(\ref{const}). This fixes $T_D$ and in turn it determines a function $m_{ \tilde{\chi}}(m_X)$. As long as the Boltzmann suppression in~(\ref{epsilonHuLXX}) is active the curve is approximately a straight line determined by $m_X/T_D\sim const\gtrsim1$. This behavior is visible in the large DM mass regime of Fig.~\ref{figHuLXX}. The Boltzmann suppression shuts off when the previous lines hit the boundary of the non-relativistic regime $m_X/T_D\sim1$. For DM masses below this point the primordial asymmetry parameter is given by $\epsilon\sim0.3$ and $m_X$ is fully determined by $r_\infty$. This effect is seen as a vertical asymptote in the low DM mass regime of Fig.~\ref{figHuLXX}.

It is evident from Figure~\ref{figHuLXX} that a large region in parameter space is associated to the asymmetric WIMP scenario $0.1\lesssim r_\infty\lesssim1$.  Here indirect signals are unsuppressed.

\subsubsection{Annihilation and direct detection}
\label{anndd}

In all the cases discussed above -- either the strongly asymmetric $r_\infty\ll1$ or strongly symmetric $r_\infty\simeq1$ limit -- we see from Fig.~\ref{fig3} that the annihilation cross section can be of thermal WIMP magnitude. Such a relatively low cross section may be induced by annihilation into SM particles mediated by non-renormalizable operators suppressed by the very same scale $\Lambda$ suppressing~(\ref{transf3}). From an effective field theory perspective in fact we expect that a generic UV completion of~(\ref{transf3}) also leads to direct couplings between the DM and the SM particles. For example, the following corrections to the superpotential and the K\"{a}hler are allowed by all the symmetries
\ba\label{Wother}
\Delta W_{sym}&=&{\lambda}^2\frac{X\overline{X}H_uH_d}{\Lambda}+\dots\\
\Delta K_{sym} &=& y^2\frac{X^\dagger XL^\dagger L}{\Lambda^2}+\dots, \label{Kother}
\ea 
and are therefore generic.% (here $\lambda$ and $y$ are $O(1)$ numbers). 

We now show that both operators~(\ref{Wother}) and~(\ref{Kother}) provide annihilation cross sections of the right strength if $\Lambda=O( \hbox{TeV})$. These operators can therefore be responsible for keeping the DM in thermal equilibrium below $T_D$, and in particular for accounting for the values of $r_\infty$ required to generate the correct relic abundance. As we will see, they can also provide direct detection signals. Let us now discuss a few annihilation modes in more detail.

The correction to the K\"{a}hler triggers the process $\psi_X\overline{\psi_X}\rightarrow l\bar l$, where $l$ are SM leptons and $\psi_X$ is a Dirac DM fermion. In order to avoid large flavor violating effects that might arise from~(\ref{Kother}), we follow~\cite{ADM} and assume that the only relevant coupling is with the third SM lepton generation. Neglecting the $\tau$ and $\nu_\tau$ masses we find~\cite{ADM}:
\ba\label{annl}
\langle\sigma_{\textrm{ann}}v\rangle=\frac{y^4}{16\pi}\frac{m_X^2}{\Lambda^4}.
\ea
It is easy to see that this channel can provide an annihilation cross section of the correct order for several DM masses. For example, for a light DM candidate, say with $m_X\sim20$ GeV one should require $\Lambda/y\lesssim200$ GeV, whereas for weak-scale DM masses, say $m_X\sim200$ GeV one sees that $\Lambda/y\lesssim700$ GeV suffices. If~(\ref{Wother}) provides the dominant annihilation mode, one finds no constraining direct detection bounds~\cite{ADM}.

The dominant annihilation mode mediated by the operator $X\overline{X}H_uH_d$, on the other hand, occurs via the s-wave annihilation either to final state fermions -- through the exchange of the (virtual) CP odd scalar $A^0$ -- or to a $h A^0$ final state. For the first channel the largest rate occurs if the final state fermions are top quarks, if kinematically allowed. Specifically, for $\overline{\psi_X}\psi_X\rightarrow {A^0}^{*}\rightarrow\bar t t$ and $ m_X > m_t$, we have
\ba\label{XXtt}
\langle\sigma_{\textrm{ann}}v\rangle=\frac{\lambda^4}{\tan^2\beta}\frac{N_c}{8\pi}\frac{m_t^2}{\Lambda^2}\frac{m_X^2}{(4m_X^2-m_{A^0}^2)^2} \sqrt{1-\frac{m_t^2}{m_X^2}}.
\ea
%\ba
%\langle\sigma_{\textrm{ann}}v\rangle=\lambda^4\frac{N_c}{8\pi}\cos^22\beta \cot^2 \beta \frac{m_t^2}{\Lambda^2}\frac{m_X^2}{(4m_X^2-m_A^2)^2} \sqrt{1-\frac{m_t^2}{m_X^2}}.
%\ea
%which is suppressed at large $\tan \beta$. 
This can be of the correct magnitude to account for the DM abundance if $A^0$ is not too heavy, $m_X$ is not too small, and $\tan\beta$ not too large. %, and $\tan \beta$ not too large. 
For example, if we take $m_X=200$ GeV $>m_{A^0}/2$ and $\tan^2\beta\approx1$ 
the averaged cross section is close to the WIMP value; for these values $\langle\sigma_{\textrm{ann}}v\rangle\approx\lambda^4\times4\times 10^{-9}$ GeV$^{-2} (1.2 \hbox{ TeV}/\Lambda)^2$. If this channel is kinematically closed the other channel may be open and provide a large enough rate. For $2m_X  > m_{h} + m_{A^0}$, we have for the s-wave annihilation $\overline{X}X \rightarrow h A^0$, 
\ba\label{XXAh}
\langle\sigma_{\textrm{ann}}v\rangle=\lambda^4 \frac{\sin^2 (\beta -\alpha) }{64 \pi \Lambda^2}
\sqrt{1-\frac{(m_{A^0}+m_{h})^2}{4 m^2_X} } \sqrt{1-\frac{(m_{A^0}-m_{h})^2}{4 m^2_X} } .
\ea
Unlike the previous cross-section, this one is not suppressed at large $\tan \beta$. 
If we take $\sin^2 (\beta -\alpha)  \approx 1$, the averaged cross-section is 
$\langle\sigma_{\textrm{ann}}v\rangle\approx\lambda^4 \times4\times 10^{-9}$ GeV$^{-2} (1.5 \hbox{ TeV}/\Lambda)^2$.  

If~(\ref{Wother}) provides the dominant annihilation mode, the DM direct detection bounds are not generally negligible. In the absence of light degrees of freedom other than the MSSM particles and the DM, the ``single nucleon'' cross section for spin-independent elastic DM scattering mediated by virtual exchanges of CP-even Higgses is 
\be 
\label{sigxn}
\sigma_{Xn} = \left(\frac{1~{\rm GeV}}{\mu_{T}}\right)^{2} \frac{\sigma_{N}}{A^{2}}, 
\ee
where the spin-independent dark matter-nucleus cross section is
\be \sigma_{N} =\frac{\mu_{T}^{2}}{\pi} \left(Z f_{p} + (A-Z) f_{n}\right)^{2} \ee
with $f_{p}$ and $f_{n}$ controlling the coupling between nucleons and the dark matter, and $\mu_{T}$ the DM-nucleus reduced mass. The nucleon coupling parameters $f_{p}$ and $f_{n}$ can be found in Appendix~\ref{appDD} and depend on the scale in Eq.~(\ref{Wother}) as well as parameters in the Higgs sector. 

Requiring that the annihilation cross section is larger than the WIMP value implies a lower bound on DM-nucleon cross section. Similarly, the bounds from direct detection can be translated into an upper bound for the annihilation cross section and thus a lower bound for $r_\infty$. 

%%%%%%%%%%%%%%%%%%%%%%%%%%%%%%%
\begin{figure}%[t] %  figure placement: here, top, bottom, or page
\begin{center}
\mbox{\subfigure{\includegraphics[width=2.8in]{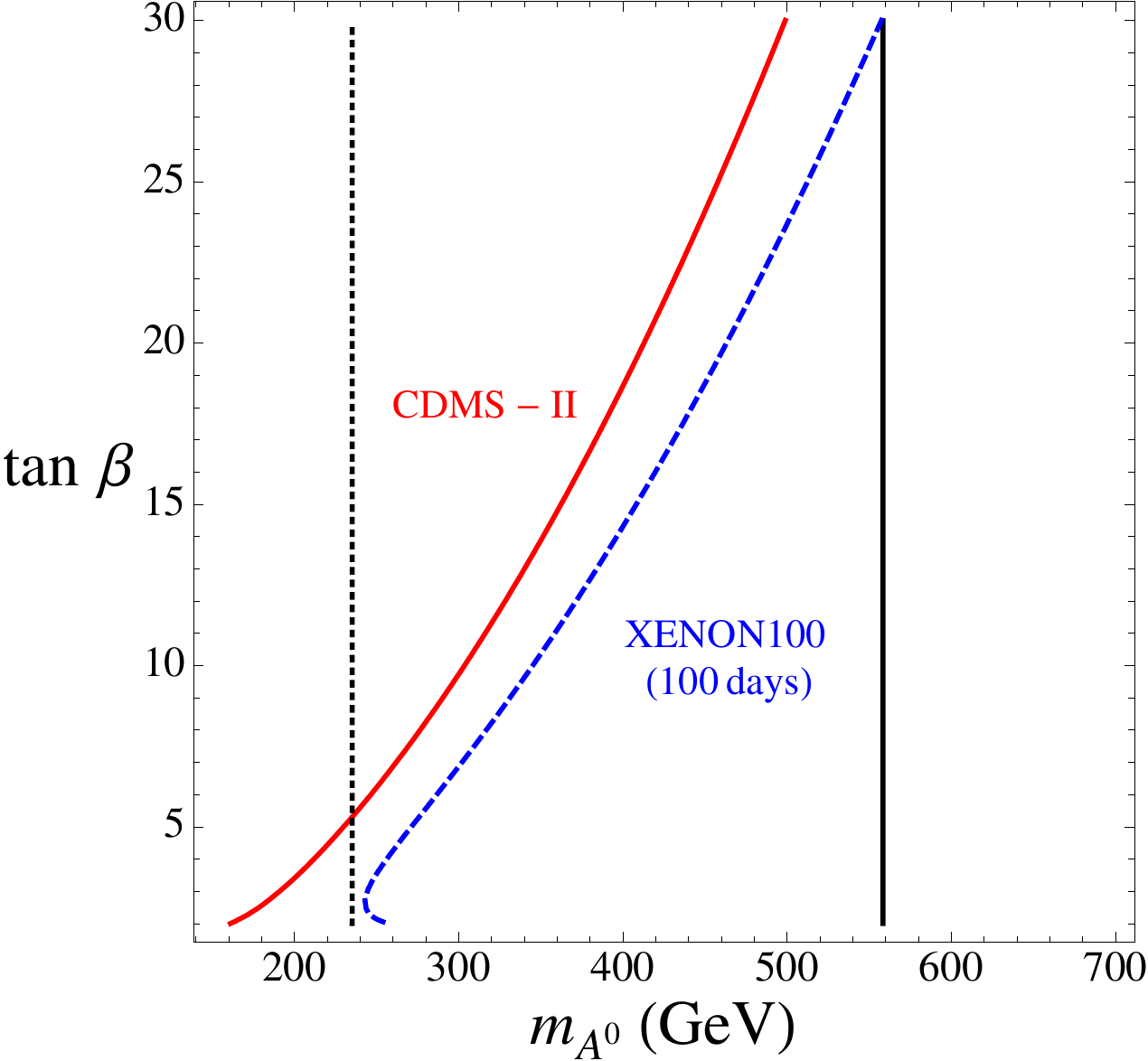}}~~~
\subfigure{ \includegraphics[width=2.8in]{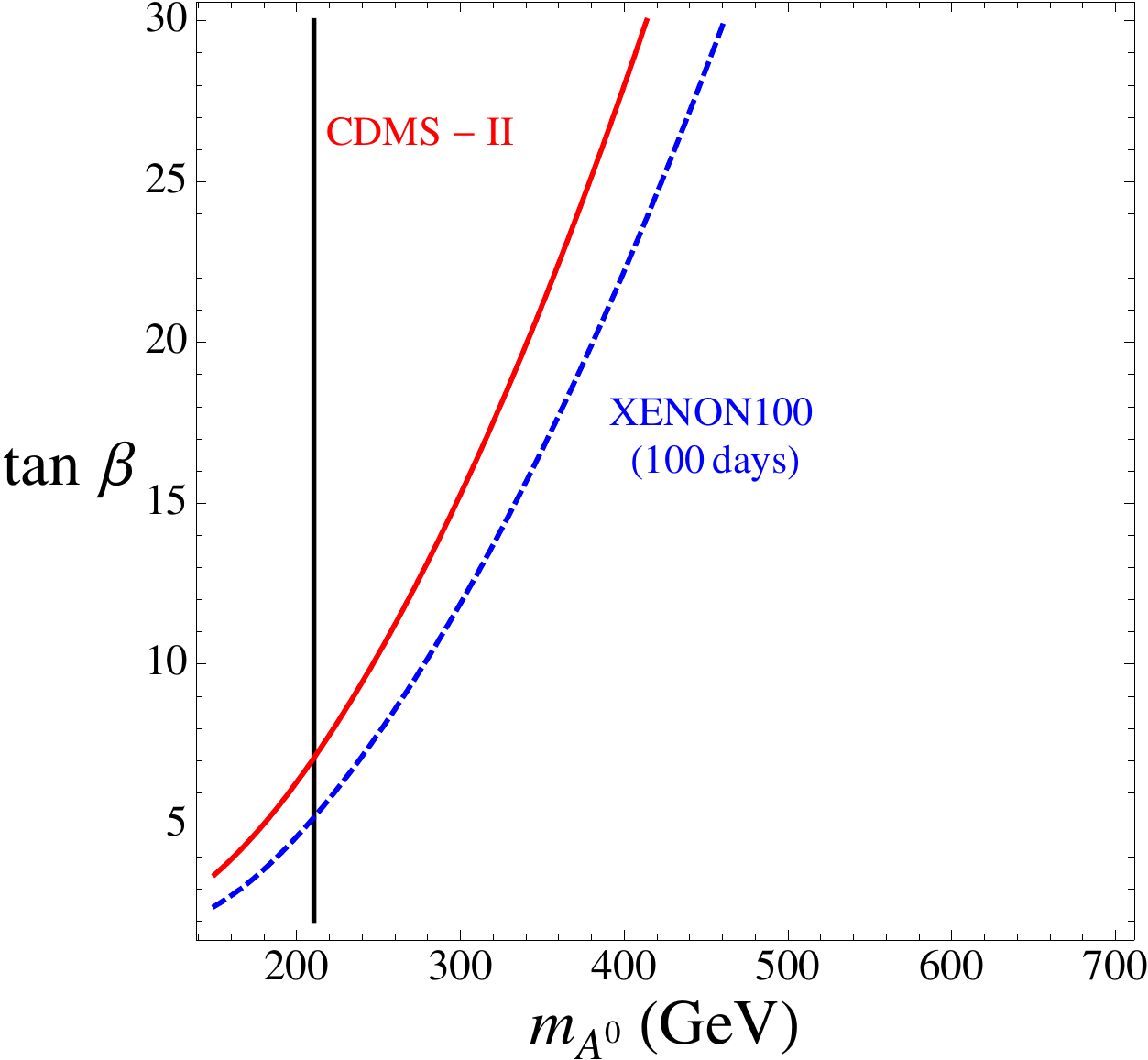}}}
\caption{\small Here we show the impact of the CDMS-II bound~\cite{cdms} $\sigma_{Xn} = 5\times 10^{-44}~\rm{cm}^{2}$ and the new XENON100 bounds~\cite{xenon} $\sigma_{Xn} = 3.3\times 10^{-44}~\rm{cm}^{2}$ on the parameter space of our model for $m_{X}=400$ GeV and $\Lambda = 700$ GeV (left plot) and $\Lambda = 1000$ GeV (right plot).  To be consistent with the null results of CDMS-II, a parameter point must lie to the right of the red curve, while to be consistent with XENON100 bounds a point must lie to the right of the dashed blue curve.  Obtaining the correct relic abundance requires  $\langle\sigma_{\textrm{ann}}v\rangle\gtrsim \sigma_{0,WIMP}$ which occurs for points to the left of the solid black curve. For reference we include the dashed black curve corresponding to $\langle\sigma_{\textrm{ann}}v\rangle= 2 ~\sigma_{0,WIMP}$, which is not present in the $\Lambda = 1000$ GeV plot.  Note that the $\Lambda = 1000$ GeV case is strongly constrained by direct detection bounds.  \label{figdd}}
\end{center}
\end{figure}
%%%%%%%%%%%%%%%%%%%%%%%%%%%%%%

To see that, let us assume that the dominant annihilation mode is $\overline{X}X \rightarrow h A^0$ and that $m_X$ is at the weak scale. We also know from Fig.~\ref{fig1} that $\langle\sigma_{\textrm{ann}}v\rangle\gtrsim \sigma_{0,WIMP} = 4.5\times10^{-9}$ GeV$^{-2} $.  The CDMS-II bound along with this constraint on the annihilation cross section is shown in Fig.~\ref{figdd}.  We also include the new XENON100 bound $\sigma_{Xn} = 3.3\times 10^{-43}~\rm{cm}^{2}$~\cite{xenon}.  In Fig.~\ref{figdd} we have taken a dark matter mass $m_{X} = 400$ GeV, a transfer operator scale $\Lambda = 700$ GeV (left plot) and $\Lambda = 1000$ GeV (right plot) and a light Higgs mass $m_{h} = 120$ GeV. We have also assumed $m_{A^{0}}^{2} \gg m_{Z}^{2}$ such that the mixing angle $\alpha$ is determined at tree-level by $\alpha \approx \beta - \pi/2$ and the heavy Higgs mass is $ m_{H}^{2} \approx m_{A^{0}}^{2}$. The relation between $\alpha$ and $\beta$ implies that the annihilation cross section contours appear as straight vertical lines.  The interested reader can find additional details in Appendix~\ref{appDD}.  %This cross section~(\ref{sigxn}) is close to the present bounds from XENON 100~\cite{xenon100} and CDMS~\cite{cdms}, and for these parameters will be completely covered by the projected bounds for XENON 100.
%\ba
%\sigma_{Xn}&=&8\,\frac{\mu_{Xn}^2m_n^2}{\sin^2(\beta-\alpha)}\langle\sigma_{\textrm{ann}}v\rangle\\\no
%&\times&\left[\frac{\cos(\beta+\alpha)}{m_{h}^2}\left(0.14\frac{\cos\alpha}{\sin\beta}-0.22\frac{\sin\alpha}{\cos\beta}\right)+\frac{\sin(\beta+\alpha)}{m_{h}^2}\left(0.14\frac{\sin\alpha}{\sin\beta}+0.22\frac{\cos\alpha}{\cos\beta}\right)\right]^2%\\\no
%&\gtrsim&3\times10^{-45}\,\,\textrm{cm}^2\,\frac{\cos^2(\beta+\alpha)}{\sin^2(\beta-\alpha)}\left(\frac{120\,\,\textrm{GeV}}{m_{h}}\right)^4\left[0.64\frac{\cos\alpha}{\sin\beta}-\frac{\sin\alpha}{\cos\beta}\right]^2
%\ea

Our analysis of the annihilation modes has been model-independent so far. We end this section by emphasizing that, once a UV completion for the operator~(\ref{transf3}) is known, alternative reactions, and constraints, might arise. As an instance, consider the UV completion proposed in~\cite{ADM}. One introduces two electro-weak doublets $D, \bar D$ with hypercharge $\mp\frac{1}{2}$ and generalized lepton number $L'=\mp1$, and two SM singlets $X, \bar X$ with $L'=\pm1$. One readily sees that integrating out the $D$ field one gets the operator~(\ref{transf3}) as well as~(\ref{Wother}) and~(\ref{Kother}). A typical feature of this UV completion is the occurrence of a mixing between the neutral component of the doublet $D$ and the DM after electro-weak symmetry breaking. This mixing induces a coupling of the DM $\psi_X$ to the $Z^0$ of order $\sim g\lambda_{u,d}v_{u,d}/\Lambda$, and thus implies new annihilation and direct detection modes mediated by a virtual gauge boson. Given the lower bound on the annihilation cross section, one finds that the direct detection signal is always too large if the mixing-induced $Z^0$ exchange is the dominant channel. One should then suppress this channel by taking $\lambda_{u,d}<O(10^{-1})$. At this point the dominant annihilation mode would be given by~(\ref{annl}). Alternatively one could consider the NMSSM, where a large contribution to~(\ref{Wother}) might be obtained integrating out the singlet $S$, also coupled to the DM via $\lambda'S\bar X X$. In this latter case the coupling of~(\ref{Wother}) would be naturally at the TeV scale, and freeze-out would thus be controlled by either $\overline{\psi_X}\psi_X\rightarrow {A^0}^{*}\rightarrow\bar t t$ or  $\overline{X}X \rightarrow h A^0$. Notice that in all of the cases discussed above the DM is assumed to be the lightest state beyond the MSSM.

\subsection{Example 2: Baryon-number violating transfer operator}

We now consider an example in which the transfer operator violates $B$ and $X$, but not $L$. The model is again an extension of the MSSM, and it is described by the following nonrenormalizable term~\cite{ADM}:
\ba\label{op}
\Delta W_{asym}=\frac{XXu^c d^c d^c}{\Lambda^2}.
\ea
The superpotential operator~(\ref{op}) preserves a global $B'=X+2B$ number, but violates $B$ and $X$ separately. The physics of this model is very similar to the lepton violating operator~(\ref{transf3}), see for example the subsection~\ref{example1}.

%A number of processes are implied by~(\ref{op}), and which of these goes out of equilibrium last is sensitive to the underlying mass spectrum. This is due to the strong temperature dependence of these reaction rates arising from Boltzmann suppression factors. The reaction rate that determines $T_D$ depends on a competition between tree and loop-level processes that differ in their Boltzmann suppression factors. 

A chemical potential analysis similar to the one presented for the $L$-violating operator in this case gives:
\ba\label{epsilonXXqqq}
\epsilon\equiv\frac{\eta_X}{\eta_B}=\left(\frac{1756 + 253 \frac{f(m/T_{sph})}{f(0)}}{5472+ 836 \frac{f(m/T_{sph})}{f(0)}- 50 \frac{f(m/T_D)}{f(0)}}\right)\frac{f(m/T_D)}{f(0)}.
\ea
Similarly to~(\ref{epsilonHuLXX}), this result is very robust and approximately reads $\epsilon\approx0.3 f(m/T_D)/f(0)$.

%As in the discussion of the $L$-violating operator we assume that the LSP is the neutralino $\tilde{\chi}$, and that $m_{\tilde{\chi}}>2m_X$. 

The typical processes transferring the asymmetry between the visible and the dark sectors
%\ba\label{XXqqq} 
%\psi_X \psi_X \leftrightarrow \overline{\psi}_X\overline{\psi}_X\,(udd)^2\quad\quad\quad \psi_X \psi_X \leftrightarrow (\overline{\psi}_X\overline{\psi}_X)^3\,(udd)^4.
%\ea 
can occur through on-shell or off-shell neutralinos or squarks. The processes involving neutralinos are always strongly suppressed by phase space, and we find that on-shell production of squarks dominates for typical masses; see Appendix~\ref{appudd} for details.  Using again our approximate expression~(\ref{Capp}) the collision term for the process $\psi_X \psi_X \rightarrow\tilde{u}\tilde{d}\tilde{d}$ is found to be
%The dominant contribution comes from
%1) the reaction $\psi_X \psi_X \rightarrow\tilde{\chi} \,udd$, with an on-shell neutralino that eventually decays with branching ratio ${\cal BR}=1$ into the $X$- and $B$-violating final state of~(\ref{XXqqq}). The former process is induced by a 1-loop diagram involving a gluino, see the Appendix for more details. Assuming comparable masses for the gluino and the squarks, I find
%\ba C\sim2.5\times10^{-8}\frac{N_c!}{(2\pi)^9}\left(\frac{g_w\alpha_s}{\Lambda^24\pi}\right)^2\frac{m_{\tilde \chi}^{14}}{m_{\tilde q}^6}\,e^{-m_{ \tilde{\chi}}/T}\left[1+O\left(\frac{T}{m_X},\frac{m_X}{m_{\tilde{\chi}}}\right)\right], \ea
%This is very suppressed. And I do not find a $T$ suppression... I am not sure this is right, then.
\ba 
C=-\frac{9\sqrt{2}\pi^{3/2}}{32\Lambda^4}\frac{N_c!}{(2\pi)^7}\,m_{\tilde q}^8\left(\frac{T}{m_{\tilde q}}\right)^{9/2}\,e^{-3m_{ \tilde{q}}/T}\left[1+O\left(\frac{T}{m_{ \tilde{q}}},\frac{m_X}{m_{\tilde{q}}}\right)\right].
\ea
where the powers of $T$ have the same origin as the ones in~(\ref{C}), and we assumed that the squarks are degenerate for simplicity. %In this case, for masses in the GeV-TeV range we have $T_D\sim(3m_{\tilde{q}}-m_X)/40$. From the condition $m_{\tilde q}>2 m_X$ we thus see that $x_D=m_X/T_D<8$. 
%The factor of $1/2^3$ refers to the branching ratio of the 3 neutralinos produced by the decay of the squarks into $B$-violating final states.

The analysis can now proceed similarly to the lepton-number violating transfer operator. The result can again be summarized with a plot analogous to Fig.~\ref{figHuLXX}, which for brevity we do not show. The comments in Section~\ref{example1} regarding s--wave annihilation and direct detection via the superpotential operator $ X \overline{X} H_u H_d$ also apply in this case.

\section{Conclusions}
\label{conc}

In this paper we studied the implication of a nontrivial DM primordial asymmetry $\eta$ on the DM relic abundance. For any $\eta$, the physical quantity controlling the effective asymmetry of the species is the fractional asymmetry $r(x)=Y^-/Y^+$. By solving the coupled Boltzmann equations for $Y^\pm$ we found that $r(x)$ depends \emph{exponentially} on both the annihilation cross section and $\eta$, see~(\ref{estimate1})~\cite{GS}. The present anti-particle population of an asymmetric DM species is therefore extremely sensitive to the strength of the dynamics. 

The parameter space in a DM model with a primordial asymmetry $\eta$ is defined as follows. The DM mass $m$ is bounded from above by the mass $m_{ADM}$ of a totally asymmetric, $r_\infty=0$, DM species, see~(\ref{ADMm}):
\ba
m\leq m_{ADM}\equiv\frac{\Omega_{DM}}{\Omega_B}\frac{\eta_B}{\eta}m_p.
\ea
\emph{For any $\eta$} one finds $m< m_{ADM}$ as soon as one departs from the condition $r_\infty=0$. In particular, very low masses $m\ll m_p$ are always allowed provided $r_\infty\rightarrow1$.

The thermally averaged annihilation cross section for an asymmetric DM candidate is instead bounded from below by the value found for a totally symmetric species of the same mass (thermal WIMP value), see~(\ref{WIMP}):
\ba
\langle\sigma_{\textrm{\small{ann}}}v\rangle\geq\langle\sigma_{\textrm{\small{ann}}}v\rangle_{WIMP}.
\ea
\emph{For any $\eta$} the cross section approaches its lowest value when $r_\infty\rightarrow1$, while a larger cross section can still be accommodated in the regime $r_\infty<1$. Notice that because of the exponential dependence of $r_\infty$ on the cross section, an effectively asymmetric DM population, say with $r_\infty\lesssim10^{-(2-3)}$, is already obtained with an annihilation cross section of just a factor $\sim(2-3)$ bigger than the thermal WIMP cross section. This latter point has interesting consequences for model-building, as it implies that new light states, often introduced in previous studies of asymmetric DM scenarios to induce large cross sections% to account for $\langle\sigma_{\textrm{\small{ann}}}v\rangle\gg\langle\sigma_{\textrm{\small{ann}}}v\rangle_{WIMP}$
, are not necessary.

A generic weakly coupled UV completion for the standard model will typically have weak scale annihilation cross sections. If the DM is stabilized by a symmetry distinguishing particles from anti-particles it is also likely that the DM has a nonzero primordial asymmetry $\eta$. As a result, asymmetric WIMP scenarios are expected to be generic. These scenarios typically lie between the extremely asymmetric and symmetric limits. Surprisingly, the indirect signals from present-day annihilation can be much larger than what one would naively expect from an asymmetric scenario.

To illustrate the generality of our conclusions we applied our results to two existing models of asymmetric DM~\cite{ADM}. We saw that the parameter space consistent with the present-day DM abundance and characterizing the asymmetric WIMP regime is sizable. This property stems from the fact that the present-day fractional asymmetry $r_\infty$ acts as new, low energy parameter. 

Moreover direct detection bounds impose interesting limits on the Higgs portal operator $\overline{X}XH_{u}H_{d}$. Although this operator was introduced to yield a sufficiently large annihilation cross section, the same interaction gives the dominant interaction with nuclei.  We found that the current bounds from CDMS-II and XENON100 constrain the model, but upcoming direct detection experiments will be sensitive to a wide range of the parameters relevant for the asymmetric WIMP through the Higgs portal. 

%As a consequence, generic UV completions of the standard model are likely to lie in the region of parameter space in which $0<r_\infty<1$, where indirect signals are still effective.  

\acknowledgments

We would like to thank Matthew McCullough for pointing out a minor error in our chemical potential analysis in the first version of this work. This work has been supported by the U.S. Department of Energy at Los 
Alamos National Laboratory under Contract No. DE-AC52-06NA25396. The preprint number for this manuscript is LA-UR 11-00565.

\appendix

\section{Collision terms with thresholds\label{CT}}

In this appendix we present a very general approximate expression for the collision term $C$ for a generic $2$- to $N$-body process with an energy threshold $M$. For definiteness we focus on $s$-channel scattering, in which the cross section $\sigma(2\rightarrow N)$ is a function of the Mandelstam variable $s$ only. Our results can be straightforwardly generalized.

In general, the collision term for a $X_1+X_2\rightarrow N$ process can be written as 
\ba
C&=&-\int \frac{d^3 p_{X_1} }{(2 \pi)^3 2 E_{X_1}} \int \frac{d^3 p_{X_2} }{(2 \pi)^3 2 E_{X_2} } f_{X_1} f_{X_2}\,\theta\left(s-M^2\right)\\\no
& \times&\Pi_f \left( \int \frac{d^3 p_f }{(2 \pi)^3 2 E_f } \right)  (2 \pi)^4 \delta^{(4)}(p_{X_1}^\mu+p_{X_2}^\mu-\sum_f p^{\mu}_f) \langle | {\cal M} |^2\rangle, 
\ea
where $E_{X_{1,2}}$ and $p_{X_{1,2}}^\mu$ are the energy and 4-momentum of the incoming particles, whereas $E_f$ and $p_f^\mu$ of the $N$ final particles, $\langle | {\cal M} |^2\rangle$ is the spin averaged matrix element, and $f_{X_{1,2}}$ are the momentum distribution functions of the $X_i$'s. The step function enforces the condition that the CM invariant energy $s$ of the incoming particles is above threshold, $s\geq M^2$. After having integrated in the final state momenta, the above formula can be equivalently expressed as an integral of the cross section times the relative velocity $\sigma(2\rightarrow N)v$ over the thermal distribution of the incoming particles. 

In our paper the energy threshold $M$ is always a heavy particle mass which by assumption it satisfies $M>2 m_X$ (see the text). In this case one can see that the incoming DM particles must be highly energetic, $p_{X_{1,2}}\simeq E_{X_{1,2}}$. Hence we can write:
\ba\label{form}
C&=&- \int \frac{d^3p_1}{(2\pi)^3}\int \frac{d^3p_2}{(2\pi)^3} \,\,\sigma\left(s/M^2\right)v\,e^{-\frac{E_1+E_2}{T}}\theta\left(s-M^2\right)\\\no
&=&-\int_{-1}^{+1} d\cos\theta \int_0^\infty dE_1\int_{M^2/[2E_1(1-\cos\theta)]}^\infty dE_2 \,\,E_1E_2\,{\cal C}\left(s/M^2\right)\,e^{-\frac{E_1+E_2}{T}},
\ea
where $s=2 E_1E_2(1-\cos\theta)$ with $\theta$ the angle between the incoming momenta, and the \emph{dimensionless} function ${\cal C}\equiv(E_1E_2v)\sigma/(2^3\pi^4)=s\sigma/(2\pi)^4$ has been defined for convenience. Notice that ${\cal C}$ is a Lorentz invariant function of $s/M^2$.

We are interested on the physics at temperatures at which the incoming particles are non-relativistic, namely $T<m_{X_{1,2}}$. Our basic assumption is therefore that the relevant temperature $T$ satisfies $M/T\gg1$. Physically, the fact that near threshold scattering dominates the collision term allows us to considerably simplify the above integral by making use of the following approximation
\ba\label{theorem}
\int_{y_{min}}^\infty dy\,\,{\cal F}(y)\,e^{-y}={\cal F}^{(n)}(y_{min})e^{-y_{min}}\,\left[1+O\left(\frac{1}{y_{min}}\right)\right],
\ea
where ${\cal F}^{(n)}$ is the lowest nonvanishing derivative of ${\cal F}$, and $y_{min}\gg1$. The theorem~(\ref{theorem}) will be used to approximate the integral in $E_2$ and in $\cos\theta$, where in our case the expansion parameter is $1/y_{min}=T/M$. Here we only sketch a few intermediate steps. Using~(\ref{theorem}) the collision term becomes, up to subleading terms in $T/M$,
\ba
C&=&-M^4\left(\frac{T}{M}\right)^{n+1}{\cal C}^{(n)}(1)\int_{-1}^{+1} dx\, [2(1-x)]^{n-1}\int_0^\infty d\varepsilon\,\varepsilon^n\,e^{-\frac{M}{T}\left(\varepsilon+\frac{1}{2(1-x)\varepsilon}\right)}\\\no
&=&-\sqrt{\pi}M^4\left(\frac{T}{M}\right)^{n+3/2}{\cal C}^{(n)}(1)\int_{-1}^{+1} dx\, [2(1-x)]^{n/2-5/4}\,e^{-\frac{M}{T}\sqrt{\frac{2}{1-x}}},
\ea
where ${\cal C}^{(n)}(1)$ is the lowest nonvanishing derivative of ${\cal C}$ at $s= M^2$, and where in the last approximate equality we have used a saddle point approximation and again neglected $O(T/M)$ terms. In the last step we make use of~(\ref{theorem}) to evaluate the integral in $x=\cos\theta$. And finally we have:
\ba\label{Capp}
C=-\sqrt{\frac{\pi}{2}}M^4\, 2^n\left(\frac{T}{M}\right)^{n+5/2}{\cal C}^{(n)}(1)\,e^{-\frac{M}{T}}\left[1+O\left(\frac{T}{M},\frac{m_{X_{1,2}}}{M}\right)\right].
\ea
Eq.~(\ref{Capp}) is the formula used in the text to estimate the collision terms for the lepton and baryon violating transfer operators. We checked that~(\ref{Capp}) is always very close to the exact, numerical result, which is consistently approached in the limit $M/T\gg1$. The dominant source of uncertainty in~(\ref{Capp}) comes from the $O(m_{X_{1,2}}/M)$ corrections.

In the general case the cross section can depend also on the Mandelstam variable $t$, in which case the function ${\cal C}$ will also depend on $\cos\theta$ explicitly, and the very last step in deriving~(\ref{Capp}) would change accordingly.

\section{Collision terms for transfer operators}
\label{colapps}
\subsection{Lepton violating transfer operator: $XXH_{u} L$}
\label{lhu}
In the early Universe the transfer operator~(\ref{transf3}) keeps $\psi_X$ in chemical equilibrium with on-shell sneutrinos ($\tilde{\nu}$) through the processes 
\ba
\psi _X \psi _X \leftrightarrow \tilde{\nu}^* 
\label{XXsnu'}
\ea 
Throughout we assume the LSP is a neutralino $\tilde{\chi}$. Then the sneutrino has a large decay rate $\tilde{\nu} \rightarrow \tilde{\chi} \nu$.
Once $H$ drops below the decay rate of the sneutrino, the processes (\ref{XXsnu'}) become Boltzmann suppressed and ineffectual. Below this scale, however, the $\psi_X$'s remain in chemical equilibrium with the neutrinos through the processes  
\ba 
\label{App:XXchi'}
 \psi_X \psi_X   \longleftrightarrow   \widetilde{\chi} \overline{\nu}
\ea
mediated by an off-shell $\tilde{\nu}$. 
These are 
described by an effective Lagrangian obtained by integrating out the sneutrino,  
\ba 
{\cal L}_{eff} \simeq  \frac{g_w}{\Lambda} \frac{v_u}{m^2_{\tilde{\nu}}} (\psi_X \psi_X) (\nu \tilde{\chi}) +h.c.
\ea 
For $m_{\tilde{\chi}} > 2 m_X$ the LSP is unstable and decays to $\psi_X \psi_X \nu$ and $\overline{\psi}_X \overline{\psi}_X  \overline{\nu}$.  Still, the processes 
\ba\label{proc}
\psi_X\psi_X \leftrightarrow \overline{\psi}_X \overline{\psi}_X \overline{\nu} \overline{\nu} 
\ea 
remain in chemical equilibrium. These reactions are generated through two types of collisions, that differ on the kinematics of the initial state particles: 
\begin{itemize} 
\item The collision is energetic enough to produce an on-shell $\tilde{\nu}$. This process is more Boltzmann suppressed compared to the following process: 
\item The collision $\psi_X \psi_X$ is energetic enough to produce a real LSP ($\tilde{\chi}$) + $\overline{\nu}$, which immediately decays back to $\overline{\psi}_X \overline{\psi}_X \overline{\nu}$.
\item The collision is below threshold and the neutralino is off-shell. 
\end{itemize} 
The last item is 
described by the effective Lagrangian \cite{ADM}
\ba 
{\cal L}_{eff} \simeq  \frac{g^2_w}{\Lambda^2} \frac{v^2_u}{m^4_{\tilde{\nu}}m_{\tilde{\chi}} }(\psi_X \psi_X)^2 \nu \nu .
\label{offshellchi}
\ea 
With the assumption that $\psi_X$ is non-relativistic at $T_D$ we find this off-shell sneutrino + off-shell neutralino process to be subdominant.  Let us now turn to approximating the collision term describing real neutralino production.

The forward and inverse processes~(\ref{proc}) contribute a ``collision term"  to the Boltzmann equation for $n_X$
\ba 
\frac{d n_X}{dt} + 3 H n_X = C[f_i]
\ea 
The ``collision term" for the $T$-invariant processes is in general 
\ba 
C[f_i] &=&- \left(\frac{1}{2}\right)^3 \left(4 \right)
 \int \frac{d^3 p_{X_1} }{(2 \pi)^3 2 E_{X_1}} \int \frac{d^3 p_{X_2} }{(2 \pi)^3 2 E_{X_2} } 
 \int \frac{d^3 p_{X_3} }{(2 \pi)^3 2 E_{X_3} }  \int \frac{d^3 p_{X_4} }{(2 \pi)^3 2 E_{X_4} }  \nonumber \\
 & & \times \int \frac{d^3 p_{\nu_1} }{(2 \pi)^3 2 E_{\nu_1} } 
  \int \frac{d^3 p_{\nu_2} }{(2 \pi)^3 2 E_{\nu_2} }
  \nonumber \\
  & & \times  (2 \pi)^4 \delta^{(4)}(\sum_i p^{\mu}_i) \langle | {\cal M} |^2 \rangle \left( f_{X_1} f_{X_2}  - f_{\overline{\nu}_1} f_{\overline{\nu}_2}  f_{\overline{X}_3} f_{\overline{X}_4} \right)
\ea 
We have assumed the occupation numbers satisfy $f_i \ll 1$.  The factor of $1/2^3$ occurs for both the forward and inverse processes and is for integrating over one eighth of phase space since the two $X$'s, two 
$\overline{X}$'s and two $\overline{\nu}$'s  are identical particles. The factor of $4$ is present because the process (\ref{proc}) changes the $X$ particle number by four units.  We note that the process $\psi_X\psi_X \leftrightarrow \psi_X \psi_X \nu \overline{\nu}$ does not contribute to the collision term since it does not change the $X$ particle number. 

If these reaction rates are large enough in the early Universe these processes are in chemical equilibrium and the collision term vanishes, 
implying the condition 
\ba 
2 \mu_X+ \mu_{\nu}=0
\ea
on the chemical potentials. 
As the Universe cools these processes eventually fall out of chemical equilibrium. When that occurs the asymmetries become frozen in.

Next we turn to showing that $C =   C_{\psi_X \psi_X \widetilde{\chi} \overline{\nu}}$ where $ C_{\psi_X \psi_X \widetilde{\chi} \overline{\nu}}$ is the collision term for the forward process $\psi_X \psi_X \rightarrow \widetilde{\chi} \overline{\nu}$.  This is expected since the neutralino does not carry $X$ number.  We will then apply our results from Appendix \ref{CT} to obtain an accurate expression for $ C_{\psi_X \psi_X \widetilde{\chi} \overline{\nu}}$.

By assumed $T-$invariance (which is true at tree-level), the amplitude for $\psi_X \psi_X  \longleftrightarrow \widetilde{\chi} 
\overline{\nu}
 \longleftrightarrow \overline{\psi}_X \overline{\psi}_X \overline{\nu}  \overline{\nu}$
factorizes as 
\ba 
{\cal M}_{\psi_X \psi_X   \overline{\psi}_X \overline{\psi}_X \overline{\nu}  \overline{\nu}} = {\cal M}_{\psi_X \psi_X \widetilde{\chi} \nu} \frac{1}{q^2 - m^2_{\widetilde{\chi}} + i m_{\widetilde{\chi}} \Gamma_{\widetilde{\chi}} } 
{\cal M}_{ \psi_X \psi_X \widetilde{\chi} \nu}
\ea 
where $q$ is the momentum-transfer along the neutralino line, $\Gamma_{\widetilde{\chi}} $ is the total decay rate of the neutralino, and ${\cal M}_{\psi_X \psi_X \widetilde{\chi} \nu}$ is the matrix element for $\psi_X \psi_X \longleftrightarrow \widetilde{\chi} \overline{\nu}$.

As noted above, the collision term for this matrix element is dominated by two regimes. In the first, $q^2 \ll m^2_{\widetilde{\chi}}$, 
and describes the off-shell production of the neutralino. It is characterized by the effective Lagrangian (\ref{offshellchi}) 
obtained by integrating out the neutralino.

The other regime corresponds to real production of the neutralino. It simplifies in the narrow-width approximation, which is well-justified since the neutralino only decays through the transfer operator. Although this regime is characterized by a large energy in the initial state to reach the kinematic threshold (and leads to a Boltzmann suppression), the rate is enhanced in the narrow-width approximation and potentially important.  

To describe the collision term for the real production of $\tilde{\chi}$, we make use of the relation, 
\ba 
\left|\frac{1}{q^2 - m^2_{\widetilde{\chi}} + i m_{\widetilde{\chi}} \Gamma_{\widetilde{\chi}} }  \right|^2 \simeq  \frac{\pi}{m_{\widetilde{\chi}} \Gamma_{\widetilde{\chi}}} 
\delta(q^2- m^2_{\widetilde{\chi}})
\label{narrowwidth}
\ea 
which is 
valid near threshold and in the narrow-width approximation.   

Then for the forward reaction, the integral over the part of final state phase space dominated by the neutralino pole simplifies to 
\ba\label{long}
C&=&-\frac{1}{2^{3}} (4)  \int \frac{d^3 p_{X_1} }{(2 \pi)^3 2 E_{X_1}} \int \frac{d^3 p_{X_2} }{(2 \pi)^3 2 E_{X_2} } f_{X_1} f_{X_2} 
 \int \frac{d^3 p_{\nu_1} }{(2 \pi)^3 2 E_{\nu_1} }  \int \frac{d^4 q}{(2 \pi)^4}  ~ \delta(q^2-m^2_{\tilde{\chi}})   \frac{\pi}{m_{\widetilde{\chi}} \Gamma_{\widetilde{\chi}}}  \nonumber \\ 
 & & \times 
  (2 \pi)^4 \delta^{(4)}(p_{X_1} + p_{X_2} -q - p_{\overline{\nu}_1} ) \langle | {\cal M}_{\psi_X \psi_X \rightarrow \widetilde{\chi} \overline{\nu}_1} |^2 \rangle
\nonumber \\
& & \times  \int \frac{d^3 p_{X_3} }{(2 \pi)^3 2 E_{X_3} }  \int \frac{d^3 p_{X_4} }{(2 \pi)^3 2 E_{X_4} }  
 \int \frac{d^3 p_{\nu_2} }{(2 \pi)^3 2 E_{\nu_2} }  \nonumber \\
&  & \times (2 \pi)^4 \delta^{(4)}(p_{\overline{\nu}_2} + p_{X_3} + p_{X_4} -q) 
\langle | {\cal M}_{\widetilde{\chi} \rightarrow  \overline{\psi}_X \overline{\psi}_X  \overline{\nu}_2} |^2 \rangle
\nonumber \\
&=& -\frac{1}{2} (4)  \int \frac{d^3 p_{X_1} }{(2 \pi)^3 2 E_{X_1}} \int \frac{d^3 p_{X_2} }{(2 \pi)^3 2 E_{X_2} } f_{X_1} f_{X_2} 
 \int \frac{d^3 p_{\nu_1} }{(2 \pi)^3 2 E_{\nu_1} }  \int \frac{d^4 q}{(2 \pi)^4}  ~ \delta(q^2-m^2_{\tilde{\chi}})   
 \frac{2 \pi \Gamma(\widetilde{\chi} \rightarrow \overline{\psi}_X \overline{\psi}_X \overline{\nu} )}{\Gamma_{\widetilde{\chi}}}  \nonumber \\ 
 & & \times 
  (2 \pi)^4 \delta^{(4)}(p_{X_1} + p_{X_2} -q - p_{\overline{\nu}_1} ) \langle | {\cal M}_{\psi_X \psi_X \rightarrow \widetilde{\chi} \overline{\nu}_1} |^2 \rangle
\nonumber \\
&=& 2~C_{\psi_X \psi_X \widetilde{\chi} \overline{\nu}} BR( \widetilde{\chi} \rightarrow \overline{\psi}_X \overline{\psi}_X \overline{\nu}) \\
   &=&  C_{\psi_X \psi_X \widetilde{\chi} \overline{\nu}}
\ea 
where $ C_{\psi_X \psi_X \widetilde{\chi} \overline{\nu}}$ is the collision term for the forward process $\psi_X \psi_X \rightarrow \widetilde{\chi} \overline{\nu}$. Namely, 
\ba 
C_{\psi_X \psi_X \widetilde{\chi} \overline{\nu}} &\equiv &  - \left(\frac{1}{2}\right) \left(2 \right)
 \int \frac{d^3 p_{X_1} }{(2 \pi)^3 2 E_{X_1}} \int \frac{d^3 p_{X_2} }{(2 \pi)^3 2 E_{X_2}}  \int \frac{d^3 p_{\overline{\nu}} }{(2 \pi)^3 2 E_{\overline{\nu}} }  \int \frac{d^3 p_{\tilde{\chi}} }{(2 \pi)^3 2 E_{\tilde{\chi}} }\nonumber \\ 
 & & (2 \pi)^4 \delta^{(4)}(\sum_i p^{\mu}_i)
 \langle | {\cal M}_{\psi_X \psi_X \widetilde{\chi} \nu} |^2 \rangle   f_{X_1} f_{X_2}
\ea
For this process the spin-averaged cross-section is 
\ba 
\langle | {\cal M}_{\psi_X \psi_X \widetilde{\chi} \nu} |^2 \rangle &=& \lambda^2 (2 p_{X_1} \cdot p_{X_2}) (2 p_{\widetilde{\chi}} \cdot p_{\overline{\nu}}) 
\\ 
&=& \lambda^2 (s -2 m^2_X)(s-m^2_{\widetilde{\chi}}) 
\ea 
where $\lambda \equiv g_w v_u/(\Lambda m^2_{\widetilde{\nu}})$ and $s\equiv Q^2 \equiv (p_{X_1}+p_{X_2})^2$. Then integral over the two-body final state phase space is trivial, since the matrix element only depends on the center-of-mass energy. Using the standard formula   
\ba 
\int d \Phi_2 (Q; p_a,p_b) = \frac{1}{(2 \pi)^5} \frac{1}{4 Q^2} \left[(Q^2-(m_a+m_b)^2)(Q^2-(m_a-m_b)^2) \right]^{1/2} 
\ea
gives 
\ba 
\int d \Phi_2(Q; p_{\overline{\nu}}, p_{\widetilde{\chi}})  \langle | {\cal M}_{\psi_X \psi_X \widetilde{\chi} \nu} |^2 \rangle (2 \pi)^4  \delta^{(4)}(\sum_i p^{\mu}_i) \nonumber \\ 
= \frac{\lambda^2}{8 \pi} (s-m^2_{\widetilde{\chi}})^{2} \left(1 -2 \frac{m^2_X}{s}\right) \theta(s - m_{\widetilde{\chi}}^2) 
\ea 
Now all but one of the angular integrals over the initial state momenta are trivial. 

The approximations we use are the following. Since $T < m_X < m_{\widetilde{\chi}}/2$, the colliding $\psi_X$'s must be highly energetic to reach the kinematic threshold. Notice that this does not contradict the statement that the $\psi_X$'s are non-relativistic at the temperature $T_D$; the previous observation is tantamount to saying that the DM particles that trigger the process of interest live far out on the Boltzmann tail. In the integrals we can therefore set $p_{X_i} = E_{X_i}$ and neglect terms of order $m^2_X/s$ and order $T/m_X$. With these approximations the collision term simplifies to 
\ba 
C_{\psi_X \psi_X \widetilde{\chi} \overline{\nu}} & = & - \int^1_{-1} d\cos \theta\int dE_1  \int dE_2\,E_1 E_2 \,e^{-(E_1+E_2)/T} \theta(s - m_{\tilde{\chi}}^2)\\\no
&\times&\frac{\lambda^2}{2^9 \pi^5}(s-m^2_{\tilde{\chi}})^2,
\ea 
with  
\ba 
s= 2 E_1 E_2 (1-\cos \theta) 
\ea 
This expression for the collision term is precisely of the form~(\ref{form}) with $M=m_{\tilde{\chi}}$ and
\ba
{\cal C}= \frac{\lambda^2}{2^9 \pi^5} m_{\tilde{\chi}}^4\left(\frac{s}{m_{\tilde{\chi}}^2}-1\right)^2.
\ea
Hence, from~(\ref{Capp}) we find ($n=2$)
\ba
C_{\psi_X \psi_X \widetilde{\chi} \overline{\nu}}=-\frac{\sqrt{2\pi}}{4}\frac{\lambda^2 m_{\tilde{\chi}}^{8}}{(2\pi)^5}\left(\frac{T}{m_{\tilde{\chi}}}\right)^{9/2}e^{-m_{\tilde{\chi}}/T}.
\ea
Finally, with~(\ref{long}) we get~(\ref{C}).

%%%%
\subsection{Baryon-violating transfer operator: $XXu^cd^cd^c$} \label{appudd}
%%%%

From the transfer operator~(\ref{op}) we consider the process
\ba 
\label{X3squark}
\psi_X \psi_X \rightarrow 3 \tilde{q} .
\ea 
This occurs at tree-level through the dimension 6 operator (\ref{op}), but its thermally averaged cross-section has a Boltzmann suppression factor that is sensitive to the mass difference between squarks and $X$. Specifically, for this process $C/(n_X H) \sim \hbox{Exp}[-(3m_{\tilde{q}}- m_X)/T]$, where we have taken equal squark masses.

The spin-averaged matrix element for this process is
\ba 
\langle |M|^2 \rangle_{spin ~avg.} &=& \frac{\,N_c!}{\Lambda^4}  (P^2-2m^2_X),
\ea 
with $P=p_{X_1}+p_{X_2}$. The 3-body phase space can be decomposed in the product of two 2-body phase space integrals using the identity
\be d \Phi_{3}(P;p_{1},p_{2}, p_{3}) = d \Phi_{2}(q;p_{1},p_{2})d \Phi_{2}(P;q, p_{3})\left(2 \pi\right)^{3} dq^{2}, ~~~q = p_{1} + p_{2}.
\ee
The collision term can be written in the general form~(\ref{form}), with
\ba
{\cal C}=\frac{N_c!}{32(2\pi)^7\Lambda^4}\int_{2m_{\tilde q}}^{P-m_{\tilde q}}dq\,\sqrt{q^2-4m_{\tilde q}^2}\sqrt{P^2-(q+m_{\tilde q})^2}\sqrt{P^2-(q-m_{\tilde q})^2}.
\ea
To make contact with the notation used in Section~\ref{CT} we stress that $P^2=s$ and $M=3m_{\tilde q}$. 

The integral expression ${\cal C}$ cannot be solved analytically. Fortunately, though, we just need its lowest nonvanishing $n$-th derivative (with respect to $s/M^2$) evaluated at threshold. One can verify that $n=2$ so that the result~(\ref{Capp}) in this case finally reads
\ba 
C_{XX3\tilde{q}}=-\frac{9\sqrt{2}\pi^{3/2}}{32\Lambda^4}\frac{N_c!}{(2\pi)^7}\,m_{\tilde q}^8\left(\frac{T}{m_{\tilde q}}\right)^{9/2}\,e^{-3m_{ \tilde{q}}/T}\left[1+O\left(\frac{T}{m_{ \tilde{q}}},\frac{m_X}{m_{\tilde{q}}}\right)\right].
\ea

There are additional processes implied by the operator $XXu^cd^cd^c$ involving the production of on-shell neutralinos. These are however less Boltzmann suppressed but receive a large additional phase space suppression.   For $m_{\tilde{q}} \gtrsim m_{\tilde{\chi}}$ we find that the process~(\ref{X3squark}) is more important. In the main body of the text we focus on this latter regime.

\section{Direct detection from the $X \overline{X} H_u H_d$ operator} 
\label{appDD}

In this section we collect standard formulae \cite{Drees-Nojiri,DMrev} relevant for the contribution of the operator 
\ba
\Delta W_{sym}= \frac{\lambda^2}{\Lambda} X \overline{X} H_u H_d 
\ea 
to direct detection. This occurs through the exchange of the two $CP$ even Higgs bosons $h$ and $H$. 

In 4-component notation one has the following interactions between the (Dirac fermion) dark matter
and the physical neutral Higgs bosons, 
\ba 
{\cal L}_{int} & = & \frac{v\lambda^2}{2 \Lambda} \overline{\psi}_X \psi_X \left[ \cos (\alpha+ \beta) h + \sin(\alpha+ \beta) H \right ] 
\ea
Next recall the couplings of the physical neutral Higgs bosons to quarks: 
\ba 
{\cal L}_{qqH/h} = \frac{m_u}{v \sin \beta} \overline{u} u \left[ h \cos \alpha + H \sin \alpha \right] +\frac{m_d}{v \cos \beta} \overline{d} d 
\left[ -h \sin \alpha + H \cos \alpha \right] 
\ea 
Integrating out the Higgses gives 
\ba 
{\cal L}_{eff} &=& \sum_q m_{q} f_q \overline{\psi}_X \psi_X  \overline{q} q 
\ea
where 
\ba 
f_u &=& \frac{\lambda^2}{2 \Lambda} \left( \frac{1}{m^2_h} \left[ \cos(\alpha+ \beta) \frac{ \cos \alpha}{\sin \beta} \right] + \frac{1}{m^2_H} \left[ \sin(\alpha+ \beta) \frac{ \sin \alpha}{\sin \beta} \right] \right) \\
&=& \frac{\lambda^2}{2 \Lambda} \left( \frac{1}{m^2_h} \left[ \cos^2(\alpha) \cot \beta- \sin \alpha \cos \alpha \right] + \frac{1}{m^2_H} \left[ \sin^2(\alpha) \cot \beta + \sin \alpha \cos \alpha  \right] \right) \no
\ea 
\ba 
f_d &=& \frac{\lambda^2}{2 \Lambda} \left( -\frac{1}{m^2_h} \left[ \cos(\alpha+ \beta) \frac{ \sin \alpha}{\cos \beta} \right] + \frac{1}{m^2_H} \left[ \sin(\alpha+ \beta) \frac{ \cos \alpha}{\cos \beta} \right] \right)  \\
&=& \frac{\lambda^2}{2 \Lambda} \left( \frac{1}{m^2_h} \left[ \sin^2(\alpha) \tan \beta- \sin \alpha \cos \alpha \right] + \frac{1}{m^2_H} \left[ \cos^2(\alpha) \tan \beta + \sin \alpha \cos \alpha  \right] \right)\no
\ea 
Next we need the quark mass operator matrix elements 
\ba 
m_i {\cal A}^{(i)}_q  \equiv \langle i| m_q \overline{q} q | i \rangle, ~~~i = p,n
\ea 
From the literature one has  \cite{Drees}:  \\
 \ba 
{\it \underline{proton}:}~~{\cal A}^{(p)}_u &\simeq& 0.023, ~~ {\cal A}^{(p)}_d \simeq 0.034, ~~ {\cal A}^{(p)}_s \simeq 0.14,~~ {\cal A}^{(p)}_{c,b,t} \simeq 0.059 \\
{\it \underline{neutron}:} ~~{\cal A}^{(n)}_u &\simeq& 0.019, ~~ {\cal A}^{(n)}_d \simeq 0.041, ~~ {\cal A}^{(n)}_s \simeq 0.14,~~ {\cal A}^{(n)}_{c,b,t} \simeq 0.059 
 \ea
Then the spin-independent target nucleus--dark-matter cross-section is 
\ba 
\sigma_N =\frac{\mu^2_T}{\pi} \left( Z f_p + (A-Z) f_n \right)^2 
\ea 
where $\mu_T$ is the reduced mass for the target+dark matter system, and 
\ba 
f_{i} = m_i \sum_q f_q {\cal A}^{(i)}_q, ~~~ i =p,n
\ea 
The ``single nucleon" cross-section quoted in experimental limits is defined to be \cite{Lewin-Smith}
\ba 
\sigma_{Xn} & \equiv & \left(\frac{1 \hbox{ GeV}}{\mu_T}\right)^2 \frac{\sigma_N}{A^2} 
\ea 
Note it is important to include the exchange of the heavy Higgs boson $H$ since: i) its couplings to down-type quarks is enhanced by a factor of $\tan \beta$ at large $\tan \beta$; and ii) in the decoupling limit where the Goldstone bosons and light physical Higgs reside in $H_u$, the heavy Higgs resides primarily in $H_d$ and its couplings to down-type quarks is proportional to $\cos \alpha \simeq 1$.

%%%%%%%%%%%%%%%%%%%%%%%%%%%%%%%%%%%%%

\bibliography{admbib}
\bibliographystyle{JHEP}

 \end{document}